\documentclass[aps,prb,twocolumn, showpacs]{revtex4}
\usepackage{amsmath}
\usepackage{amssymb}
\usepackage{bm}
\usepackage{epsfig}
\usepackage{graphicx}
\usepackage{color}
\newcommand{\hF}{\mathop{{}_2\rm{F}_1}}
\newcommand{\grad}{\mathop{\rm grad}}
\newcommand{\divv}{\mathop{\rm div}}
\renewcommand{\Re}{\mathop{\rm Re}}
\renewcommand{\Im}{\mathop{\rm Im}}

\def\be{\begin{equation}}
\def\ee{\end{equation}}
\def\bea{\begin{eqnarray}}
\def\eea{\end{eqnarray}}

\begin{document}

\newcount\timehh  \newcount\timemm
\timehh=\time \divide\timehh by 60
\timemm=\time
\count255=\timehh\multiply\count255 by -60 \advance\timemm by \count255

\title{Pump-Probe Faraday Rotation and Ellipticity in an Ensemble of Singly
Charged Quantum Dots}
\author{I. A.~Yugova}
\affiliation{Institute of Physics, St. Petersburg State University, 198504 St.-Petersburg, Russia}

\author{M. M.~Glazov}
\affiliation{Ioffe Physical-Technical Institute RAS, 194021 St.-Petersburg, Russia}

\author{E. L.~Ivchenko}
\affiliation{Ioffe Physical-Technical Institute RAS, 194021 St.-Petersburg, Russia}

\author{Al. L.~Efros}
\affiliation{Naval Research Laboratory, Washington DC 20375, USA }

\begin{abstract}
A description of spin Faraday rotation, Kerr rotation and
ellipticity signals for single- and multi-layer ensembles of
singly charged quantum dots (QDs) is developed. The microscopic
theory considers both the single pump-pulse excitation and the
effect of a train of such pulses, which in the case of long
resident-electron spin coherence time leads to a stationary
distribution of the electron spin polarization. {The calculations
performed for single-color and two-color pump-probe setups show
that the three {experimental} techniques{: Faraday 
rotation, Kerr rotation and ellipticity measurements} provide 
{complementary}
information about an inhomogeneous ensemble of QDs.}  The
microscopic theory developed for a three-dimensional ensemble of
QDs is shown to agree with the phenomenological description of
these effects. The typical time-dependent traces of pump-probe
Faraday rotation, Kerr rotation  and ellipticity signals are
calculated for various experimental conditions. 
\end{abstract}
\date{\today} %paper-June18ae.tex, printing time = \number\timehh\,:\,\ifnum\timemm<10 0\fi \number\timemm}

\maketitle

\newpage

\section{Introduction}
It is impossible to overestimate the role, which pump-probe spin
Faraday and Kerr rotation measurements have played and continue to
play in developing of spintronics, a new area of science and
technology that tries to utilize an electron spin, in addition to
its charge, in various semiconductor devices
\cite{Wolf_Science01,Spintronics_02}. The discovery of a very long
spin coherence time in bulk GaAs \cite{Kikkawa_PRL98} %(see also\cite{korenev}) 
and II-VI compound quantum wells
\cite{Kikkawa_Science97} using these highly sensitive techniques
was one of the cornerstones for the initiation of spintronics, and
today pump-probe spin-dependent spectroscopy has become a
common way to study carrier spin coherence in bulk crystals
\cite{GaN,Kennedy_PRB06}, quantum wells (QWs)
\cite{Zhukov_PRB07,QW} and quantum dot (QD) samples
\cite{Gupta_PRB99,Petroff_APL01,Greilich_Science06,Greilich_PRL06,Greilich_PRB07}.
At the same time Kerr rotation measurements have become the
most visual and impressive method to study electron spin
transport \cite{Kikkawa_Nature99,Crooker_PRL05}, spin accumulation
and injection \cite{Stephens_PRL04,Crooker_Science05}, and
the spin-Hall effect \cite{Kato_Science04,Sih_PRL06}.

The schematic illustration of the pump-probe measurement
techniques is shown in Fig.~1. A first short intense pulse of
circularly polarized light (a pump pulse) generates the
nonequilibrium spin-oriented electrons and holes and creates a
macroscopic spin polarization~\cite{Opt_orient}. In a constant
transverse magnetic field, $\bm B$, applied to the sample the
macroscopic polarization starts to precess around the field
direction. On the microscopic single-spin level, the precession is
connected with a coherent superposition of two spin levels split
by the magnetic field. The superposition is created by a short
pulse of circularly polarized light and  the quantum mechanical
beating of this coherent superposition occurs at the Larmor
precession frequency of the applied magnetic field, $\Omega_{\rm L} = g_e\mu_B B/ \hbar$, where $g_e$ is the electron $g$-factor,
and $\mu_B$ is the Bohr magneton. The optically created
polarization and its precession can be probed by short pulses of
linearly polarized light via rotation of their polarization plane
after the propagation through the photoexited medium (Faraday
effect) or reflection from this medium (Kerr effect). The short
probe pulses of linearly polarized light show a remarkable
sensitivity to the practically instant population of electron and
hole spin sublevels. The pump-probe techniques allow one to study spin
dynamics of resident carriers that are also polarized by the pump
pulse during their coherence time, which exceeds the typical
photoluminescence decay time by several orders of magnitude. These
advantages make the pump-probe Faraday and Kerr rotation
techniques suggested more than 15 years ago
\cite{Awschalom_PRL85,Baumberg_PRL94,Baumberg_PRB94,Harley_SolStatCommun94}
to be a powerful tool to study the carrier spin dynamics.

\begin{figure}[hptb]
\includegraphics[width=\linewidth]{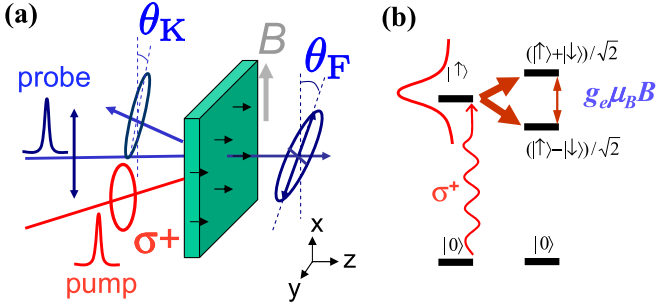}
\caption{Schematic illustration of (a) the pump-probe Faraday
and Kerr rotation measurement techniques and (b) a coherent superposition
created by a short pulse of $\sigma^+$ polarized light from the two spin
states split in a transverse magnetic field, $\bm B$. $\theta_{\rm F}$ and
$\theta_{\rm K}$ are the Faraday and Kerr rotation angles, respectively.}
\label{fig:scheme}
\end{figure}

The pump-probe spin-dependent rotation techniques are especially
useful for manipulation and measurement on electron spin
polarization in singly charged QDs due to a very long coherence
time of resident-electron spins. This time in QDs could be as long as several
microseconds~\cite{Petta_Science05,Greilich_Science06} and it exceeds the spin coherence time measured in bulk GaAs \cite{korenev,Kikkawa_PRL98} by two orders of magnitude. The
resonant short pulse excitation of such QDs by circularly
polarized light leads to practically deterministic creation of an
electron spin polarization \cite{Greilich_PRL06}. A train of such
pump pulses results in complete synchronization of electron spins
if their precession frequency satisfies the phase synchronization
condition (PSC), which is fulfilled when the train repetition
period being shorter than the single electron spin coherence time
is equal to an integer number of the Larmor precession periods
\cite{Shabaev_PRB03}. This synchronization leads to the mode
locking of electron spin coherence in an ensemble of QDs
\cite{Greilich_Science06}, the effect which allows one to overcome
the dephasing of electron spin polarization connected with a
dispersion of precession frequencies and to control the ensemble
polarization during a single electron spin coherence time. The
mode-locking effect in self-organized {(In,Ga)}As/GaAs QDs is enhanced by
the nuclear induced frequency focusing effect, which shifts the
electron spin precession frequencies of the entire ensemble of QDs
to the modes satisfying the PSC \cite{Greilich_Science07}. This
last phenomenon opens exciting opportunities to create an entire
ensemble of QDs with more than one million spins having a single
precession frequency \cite{Greilich_09SM} and their controlled
manipulation by short pulses of various polarizations
\cite{Greilich_09Man,Economou_PRL07}.

Despite the tremendous success of the pump-probe Faraday and Kerr
rotation measurement techniques, their microscopic theoretical
descriptions for QD structures are absent to the best of our
knowledge. For bulk semiconductors and QWs  the theory of
magneto-optical photoinduced Faraday effect was developed,
respectively, by Aronov and Ivchenko \cite{Aronov_Ivchenko} and by
the authors of Refs.
\cite{Zhukov_PRB07,Ostreich_PRL95,Linder_PhysE98}. In the present
paper we develop such a theory for the array of QDs considered as
an ensemble of independent localized oscillating dipoles. This
approximation, which generally imposes a restriction on the QD concentration,
is usually sufficiently accurate in self-organized QD samples like those used,
e.g., in Ref. \cite{Greilich_PRL06}.

In the standard pump-probe Faraday rotation experiments, the
transmitted probe light is split into two linearly polarized beams
with orthogonal polarizations oriented at $\pm 45^\circ$ angles
relative to the initial light polarization (or the polarization of
the probe light transmitted through the unpumped sample). Then the difference of the time-integrated intensities 
of the split beams is measured as a
function of the delay between pump and probe pulses
\cite{Crooker_PRB97}. To describe the experimental setup
 we introduce two pairs of axes, $x, y$ and $x', y'$, rotated by  a $45^\circ$ angle  with
respect to each other. The initial polarization of the probe light is along $x$ axis. This allows us to define the
experimentally measured spin Faraday signal, $\mathcal F$, as
\begin{equation}
 \label{eq:1}
\mathcal F = \lim_{z\to +\infty} \int_{0}^{T_{\rm exp}} 
\left[|E^{(t)}_{x'}(z,t)|^2 - |E^{(t)}_{y'}(z,t)|^2\right] \mathrm dt\:,
\end{equation}
where $E^{(t)}_{x'}(z,t)$ and $E^{(t)}_{y'}(z,t)$ are respectively the $x'$- and $y'$-components of an electric field
of the transmitted probe
light  at time $t$. They are connected with the $x$- and
$y$-components of the electric field: $E_{x'}=(E_{x} - E_y)/\sqrt{2}$ and $E_{y'}=(
E_{x}+E_y)/\sqrt{2}$. Equation \eqref{eq:1} assumes the probe
light source to be positioned at $z \to -\infty$. 

In this form Eq.~\eqref{eq:1} is derived for the case where the sample is subject to a periodic train of pump and probe pulses repeated with a certain period $T_R$. {The integration in Eq.~\eqref{eq:1} takes place over the measurement time, $T_{\rm exp}$, which exceeds by far all other time constants in the experiment, such as spin precession and relaxation times and pulse repetition period.} For the case of a single pump and single probe pulse the integration in Eq.~\eqref{eq:1} is effectively carried out during the probe pulse duration, $\tau_p$.

The Kerr effect
is measured in the reflection geometry and its magnitude is defined as
\begin{equation}
 \label{eq:2}
\mathcal K = \lim_{z\to -\infty} \int_{0}^{T_{\rm exp}} 
\left[|E^{(r)}_{x'}(z,t)|^2 - |E^{(r)}_{y'}(z,t)|^2\right] \mathrm dt\:,
\end{equation}
where $E^{(r)}_{x'}(z,t)$ and $E^{(r)}_{y'}(z,t)$ are respectively the $x'$- and $y'$-components of the reflected probe pulse electric field. The
probe-pulse rotation angles in the Faraday and Kerr rotation
measurements can approximately be expressed in a simple form
\cite{Magnetooptics}
\begin{equation} \label{angles}
\theta_{\rm F} \approx \mathcal F/2\mathcal I\:, \quad \theta_{\rm K} \approx \mathcal K/2\mathcal I\:,
\end{equation}
if $|\theta_{\rm F,K}| \ll 1$. Here $\mathcal I$ is the total
time-integrated intensity of {the transmitted or reflected probe pulse, respectively.}

The pump-probe ellipticity measurement is another way to study
spin dynamics in samples photoexcited by the polarized pump pulse
\cite{Dutt_PRB06}. {The experimental setup in this case is similar to setup used for the Faraday rotation measurements but with 1/4 waveplate.} The ellipticity signal in transmission is
defined by the following integrated difference
\begin{equation}
 \label{eq:3}
\mathcal E = \lim_{z\to +\infty}{\int_{0}^{T_{\rm exp}} } \left[|E^{(t)}_{\sigma^-}(z,t)|^2 -
|E^{(t)}_{\sigma^+}(z,t)|^2\right] \mathrm dt\:,
\end{equation}
where $E^{(t)}_{\sigma^\pm}=(E_x^{(t)} \mp \mathrm
iE_y^{(t)})/\sqrt{2}$ are the circular $\sigma^+$ and $\sigma^-$
components of the transmitted probe pulse. For small ellipticity
values, the so-called angle of ellipticity, $\varepsilon$, is
given by \be \varepsilon \approx \mathcal E/2\mathcal I\:.
\label{eg:4} \ee

In this paper we calculate microscopically the magnitudes of the
single- and two-color pump-probe Faraday and Kerr rotation
signals, ${\mathcal F}$ and ${\mathcal K}$, and the ellipticity
signal ${\mathcal E}$ in an ensemble of singly charged QDs
resonantly excited by a single short light pulse of an arbitrary
shape or {an infinite} train of such pulses. The calculations show that the
three measurement techniques explore spin polarization properties
of different subsets of QDs due to inhomogeneous broadening of the
resonant transition energies in the QD ensemble, and the results
of these measurements may be nonidentical. As a result, the
time-dependent traces measured by these three techniques
significantly differ from each other and are very sensitive
functions of the pump and probe excitation frequencies, their
detuning, and the dependence of the electron $g$-factor and the
oscillator transition strength on the trion excitation
frequencies. The traces of the Kerr rotation signal depend also on
the thickness of the cap layer. The average electron spin
precession frequencies measured in these experiments can differ as
well.

Our paper is organized as follows. In Sec.~\ref{sec:pump} we
provide a theoretical description of electron spin polarization
created in an ensemble of QDs by a single short pulse of circularly
polarized light or by an infinite train of such pulses. The general
microscopic theory of probing this spin polarization in the QD
ensemble is presented in Sec.~\ref{sec:probe}.  For a
three-dimensional ensemble of QDs we compare the developed
approach with the standard phenomenological description of
magneto-optical pump-probe effects within the effective-medium
approximation. In the final Sec. IV we calculate typical
time-dependent traces of the two color pump-probe Faraday and Kerr
rotation and the ellipticity signals and briefly compare the obtained
results with available experimental data.

\section{Creation of electron spin polarization in singly charged quantum dots}\label{sec:pump}

In what follows we consider a planar array of singly charged zinc blende based QDs
grown along the axis $z \parallel [001]$. The QDs are
self-organized during molecular beam epitaxy growth on the wetting
layer. The lateral size remarkably exceeds their height, which
serves a quantization axis for the electron $\pm 1/2$
 spin states and heavy-hole $\pm 3/2$ spin states  responsible for the
dominating optical transitions. In the absence of a magnetic
field, the ground state of a singly charged QD is two-fold spin
degenerate. The first excited state of such a QD under interband
transitions is a singlet trion, which consists of two electrons
occupying the same size-quantized level with opposite spins and a
heavy hole in one of the two degenerate  states: $\pm 3/2$. The
optical selection rules for the resonant excitation of these trion
states and their radiative decay are very restrictive. The
$+3/2$ trion states can be created only by the $\sigma^+$
circularly polarized light and only in a QD where the resident
electron has the spin projection $+1/2$. These $+3/2$ trion states
can radiatively decay only into the initial $+1/2$ spin states. At
the same time $\sigma^+$ circularly polarized light does not
affect an electron with the spin projection $-1/2$. The same rules
with the sign reversal ``+'' $\leftrightarrow$ ``$-$'' are applied
for the optical excitation of the $-3/2$ trion state. It is
important to notice that the singlet trion in these QDs does not
have an optical transition dipole component along the $z$ axis.

\subsection{Pumping of electron spins in quantum dots}

Let us first consider the effect of QD photoexcitation by a short
electromagnetic pulse with the carrier frequency
$\omega_{ \mbox{}_{\rm P}}$ close to the trion resonant
frequency $\omega_0$. We also assume that the pulse duration time
$\tau_p$ is short compared with other times: the
spin relaxation times of a resident electron and a photohole
forming the trion; the trion radiative lifetime; and the spin
precession period of an electron and a heavy hole in an
external magnetic field. According to the selection rules the
interaction of the QD with the electromagnetic wave is described
by the Hamiltonian
\begin{equation}
 \label{ham:circ}
\hat V(t) = -\int [\hat d_+(\bm r) E_{\sigma^+}(\bm r,t) +
\hat d_-(\bm r)E_{\sigma^-}(\bm r,t)] \mathrm d^3 r\:,
\end{equation}
where  $\hat d_\pm(\bm r)=[\hat d_x(\bm r) \pm \mathrm i \hat
d_y(\bm r)]/\sqrt{2}$ are the circularly polarized components of
the dipole moment density operator, and $E_{\sigma ^\pm}(\bm r,t)$
are the circularly polarized components of the electric field of a
quasi monochromatic electromagnetic wave. {The electric field of this wave is} defined as \be
 \bm E(\bm r, t) = E_{\sigma^+}(\bm r,t) \bm o_+  + E_{\sigma^-}(\bm r,t)\bm o_- + {\rm c.c.}\:,
\label{elfield} \ee where $\bm o_\pm$ are the circularly polarized
unit vectors related to the unit vectors ${\bm o}_x \parallel x$
and ${\bm o}_y \parallel y$ by $\bm o_\pm = (\bm o_x \pm \mathrm i
\bm o_y)/\sqrt{2}$. Here the both components $E_{\sigma ^+}$ and $E_{\sigma^-}$  are proportional
to the exponential function $ \mathrm e^{-\mathrm i \omega_{
\mbox{}_{\rm P}} t}$. 

The incident electromagnetic field induces optical transitions
between the electron state and the trion state creating a coherent
superposition of them. In accordance with the selection rules the
$\sigma^+$ circularly polarized light creates a superposition of
the $+1/2$ electron and $+3/2$ trion states while the
$\sigma^-$ polarized light creates a superposition of the $-1/2$
electron and $-3/2$ trion states. In order to describe these
superpositions it is convenient to introduce a four component
wavefunction
\begin{equation}
 \label{wave}
\Psi = \left(
\psi_{1/2},
\psi_{-1/2},
\psi_{3/2},
\psi_{-3/2}
\right)\:,
\end{equation}
where the $\pm 1/2$ subscripts denote the electron spin projection
and $\pm 3/2$ refer to the spin projection of a hole in the trion.
The electron spin polarization is expressed in terms of $\psi_{\pm
1/2}$ as follows 
\bea
S_z&=&\left(|\psi_{1/2}|^2-|\psi_{-1/2}|^2\right)/2\:,\nonumber\\
S_x&=&\Re(\psi_{1/2}\psi_{-1/2}^*)\:,
~S_y=-\Im(\psi_{1/2}\psi_{-1/2}^*)\:. \label{eq:12} 
\eea
Hereafter
we completely neglect all other excited states of a QD, e.g.,
triplet trion states, and treat the QD optical excitation within
the four-level model. In this approximation the action of a short
pulse on the charged QD can be described by the following
equations
\begin{eqnarray}
 \label{system:1}
&&\mathrm i \hbar \dot{\psi}_{3/2} =\hbar \omega_0 \psi_{3/2} + V_+(t) \psi_{1/2} \:,
\\ &&\mathrm i \hbar \dot{\psi}_{1/2}= V_+^*(t) \psi_{3/2}\:,
\nonumber
\\  \label{system:2}
&&\mathrm i \hbar \dot{\psi}_{-3/2} = \hbar \omega_0 \psi_{-3/2} + V_-(t) \psi_{-1/2}
\:,\\
&&\mathrm i \hbar \dot{\psi}_{-1/2} = V_-^*(t) \psi_{-3/2}\:.\nonumber
\end{eqnarray}
Here $\dot{\psi} \equiv \partial \psi/\partial t$ and the
time-dependent matrix elements $V_\pm (t) = -\int \mathsf d(\bm
r)E_{\sigma^\pm}(\bm r,t)\mathrm d^3 r$ describe the light
interaction with a QD. The strength of this interaction is
characterized  by the effective transition
dipole~\cite{ivchenko05a}
\bea
\label{dpm}
\mathsf d(\bm r) &= &\langle 1/2 |\hat d_- (\bm r)|3/2\rangle =
\langle - 1/2 |\hat d_+ (\bm r)|-3/2 \rangle \nonumber \\
&=&-\mathrm i
\frac{ e  p_{cv}}{\omega_0 m_0}{\mathsf F}(\bm r, \bm r)~,
\eea
which is the matrix element of the operators $\hat{d}_{\pm}({\bm
r})$ in Eq.~\eqref{ham:circ} calculated between the wave functions
of the valence band, $|\pm 3/2\rangle$, and the conduction band,
$|\pm 1/2\rangle$, all taken in the electron representation. In
Eq. \eqref{dpm}, $e$ is the electron charge, $m_0$ is the free
electron mass, and $p_{cv} = \langle \mathsf S|\hat p_x| \mathsf
X\rangle=\langle \mathsf S|\hat p_y| \mathsf Y\rangle =\langle
\mathsf S|\hat p_z| \mathsf Z\rangle$ is the interband matrix
element of the momentum operator taken between the conduction- and
valence-band Bloch functions at the $\Gamma$ point of the
Brillouin zone, $\mathsf S$ and $(\mathsf X$, $\mathsf Y$,
$\mathsf Z)$, respectively. Finally,~\cite{esser} 
\begin{equation}
\label{psi:rr}
{\mathsf F}({\bm r}_e, {\bm r}_h) = \varphi_h(\bm r_h) \varphi_{e}^{({\rm tr})}(\bm r_e)\int \mathrm d^3r' \varphi_e(\bm r')\varphi_e^{({\rm tr})}(\bm r'),
\end{equation}
where $\varphi_e^{({\rm tr})}$ and $\varphi_h$ are, respectively, the electron and heavy-hole envelope functions in a trion, and $\varphi_e$ is the envelope function of a single (resident) electron confined in a QD. These wave functions are chosen to be real. In derivation of Eq.~\eqref{psi:rr} we assumed that the QD is small enough so that the electron and hole motion in a trion can be treated independently, but allowed for the different orbital functions of the resident electron and the electrons in a singlet trion.~\cite{semina08} Note, that 
in Eq.~\eqref{psi:rr} $\mathsf F({\bm r}_e, {\bm r}_h)$ is taken at the coinciding
coordinates of the carriers, ${\bm r}_e = {\bm r}_h = {\bm r}$.

Due to a very short timescale of the pump pulse  we completely
neglect the electron and hole spin precession {during pulse action} as well as  spin
dephasing and radiative decay processes in Eqs. (\ref{system:1})
and (\ref{system:2}). The dynamics of the electron spin
polarization described by these equations was considered before in
Ref.~\cite{Greilich_PRL06} in the case of a rectangular shape
pulse with the resonant carrying frequency $\omega_{ \mbox{}_{\rm
P}}=\omega_0$. This paper generalizes the consideration for
detuned pulses of an arbitrary shape.

To be specific we consider the excitation of the QD by a
$\sigma^+$ polarized light pulse; the difference in the dynamics
of the electron spin polarization created under the $\sigma^-$
photoexcitation is briefly discussed below. Before the pump pulse
arrival the QD is always in the ground state because the pump repetition period is much longer than the trion lifetime in the QD. For the $\sigma^+$
polarized excitation the component $\psi_{-1/2}$ of the QD
wavefunction is conserved, and $\psi_{-3/2} \equiv 0$. This allows
one to reduce the set of Eqs.~\eqref{system:1} to a single
equation for the component $\psi_{1/2}(t)$
\begin{equation} \label{single}
\ddot{\psi}_{1/2} - \left( \mathrm i \omega'+ \frac{\dot{f}(t)}{f(t)}
\right) \dot{\psi}_{1/2} + f^2(t) \psi_{1/2} =0\:.
\end{equation}
Here $\omega' = \omega_{ \mbox{}_{\rm P}} - \omega_0$ is the
detuning between the pump frequency and the trion
resonance frequency, and $f(t)$ is a smooth envelope of pump pulse
defined as
\[
f(t) = -\frac{\mathrm e^{\mathrm i \omega_{\mbox{}_{\rm P}} t}}{\hbar}\int \mathsf d(\bm r)
E_{\sigma_+}(\bm r,t)\mathrm d^3 r\:.
\]

It follows from Eq. \eqref{single} that the values $\psi_{1/2}$
and $\dot{\psi}_{1/2}$ before and after the pulse action are
connected linearly \cite{arnold}. Taking into account that the
initial conditions for Eqs.~\eqref{system:1} are
$\psi_{1/2}(-\infty) ={\rm const}$, $\psi_{3/2}(-\infty) =0$ and,
therefore, $\dot{\psi}_{1/2}(-\infty) = 0$, one can represent the
solution of Eq.~\eqref{single} at $t \gg \tau_p$, i.e., after the
pulse is over, as
\begin{equation} \label{an}
\psi_{1/2}(\infty) = Q \mathrm e^{\mathrm i \Phi}\psi_{1/2}(-\infty)\:.
\end{equation}
Here the real coefficient $Q$ satisfies the condition $0 \leqslant
Q \leqslant 1$ and the phase $\Phi$ can be chosen in the interval
between $- \pi$ and $\pi$. Both parameters are determined by
the pump pulse shape, power and detuning. Equations (\ref{eq:12})
and (\ref{an}) determine the modification of the
electron spin polarization from a  short pulse of an arbitrary
shape. Taking into account that $\psi_{-1/2}$ is conserved under
the $\sigma^+$ polarized excitation, the electron spin before the
pulse arrival, $\bm S^- = (S_x^-,S_y^-,S_z^-)$, and just after the
end of the pulse, $\bm S^+ = (S_x^+,S_y^+,S_z^+)$, are connected
by
\begin{subequations} \label{pm}
\begin{eqnarray}
S_z^+ &=&   \frac{Q^2-1}{4} +\frac{Q^2+1}{2}S_z^-\:, \label{szpm}\\
S_x^+ &=& Q\cos{\Phi} S_x^- +Q\sin{\Phi} S_y^-\:,\label{sxpm} \\
S_y^+ &=& Q\cos{\Phi} S_y^- - Q\sin{\Phi} S_x^-\:.\label{sypm}
\end{eqnarray}
\end{subequations}
Although Eqs.~\eqref{pm} are derived for the pure spin states they
are valid as well for the mixed states whenever the pulse duration
$\tau_p$ is much shorter than the spin relaxation times in the QD
and the time of electron and hole spin precession in a transverse
magnetic field.

Using Eqs.~\eqref{system:1} one can show that $|\psi_{1/2}(t)|^2 +
|\psi_{3/2}(t)|^2 = |\psi_{1/2}(-\infty)|^2$. It follows then that
the $z$ component of the post-pulse trion spin polarization
formally defined as $J_z = (|\psi_{3/2}(\infty)|^2 -
|\psi_{-3/2}(\infty)|^2)/2$ is equal to
\begin{equation} \label{Jz}
J_z = S_z^- - S_z^+\:.
\end{equation}

Derivation of the relation between $\bm S^+$ and $\bm S^-$
established by the $\sigma^-$ circularly-polarized pulse gives
equations similar to Eqs. (\ref{pm}).  One has, however, to change
the sign of the first term in Eq. (\ref{szpm}) and replace $\Phi$
by $- \Phi$ in Eqs. (\ref{sxpm}) and (\ref{sypm}). 

Equations~\eqref{pm} and \eqref{Jz} are the main result of this
Section. They show how the short pulses of circularly polarized
light create and control the electron spin polarization in
$n$-type QDs under the resonant trion excitation. At low pump
intensities the electron spin is weakly affected and a value of $Q
\mathrm e^{\mathrm i \Phi}$ slightly deviates from unity. For high
intensities the coefficient $Q$ noticeably decreases and the phase
$\Phi$ shifts from zero. One can see from Eq.~\eqref{szpm} that
the $\sigma^+$ circularly polarized pulse modifies the
$z$-component of electron spin polarization by
$S_z^+-S_z^-=(Q^2-1)(1 + 2 S_z^-)/4$. The pump pulse
also leads to the in-plane rotation of the electron spin
polarization along the light propagation direction for pulses with $\Phi\neq
0$ (see
Eqs.~\eqref{sxpm} and \eqref{sypm}) similarly to spin rotation in a longitudinal magnetic field
${\bm B} \parallel z$.

It is easy to show that the phase $\Phi$ becomes nonzero due to
detuning between the resonant and pump frequencies.
Indeed, under resonant conditions, $\omega' = 0$, or small
detuning, $\omega' \tau_p \ll 1$, one can neglect $\mathrm i \omega'$  as compared with
$\dot f(t)/f(t) \sim \tau_p^{-1}$ in
Eq.~\eqref{single}  and produce the solution in the
form
\be \psi_{1/2}(t) = \psi_{1/2}(-\infty)
\cos{\left[\int_{-\infty}^t f(t')\mathrm dt'\right]}. \ee 
The
direct comparison with Eq. \eqref{an} gives $\Phi\equiv 0$ and $Q
= \cos{(\Theta/2)}$, where \be \Theta = 2\int_{-\infty}^\infty
f(t')\mathrm dt' \ee is the effective pulse area. This
consideration shows that only detuned pump pulses give rise to
$\Phi \neq 0$ and generate an effective magnetic field acting on
an electron spin in the QD. The pulse tuned resonantly to the
trion transition causes no rotation of the in-plane spin
components, and the electron spin dynamics is independent of the pulse
shape and is controlled only by the pulse area $\Theta$.

Let us now analyze the dependence of $Q$ and $\Phi$ on the pulse
parameters for the detuned pulses. In the general case of an
arbitrary pump pulse power, its arbitrary detuning and shape,
Eq.~\eqref{single} can be solved only numerically. The analytical
solutions of Eq.~\eqref{single} are known for the two pulse
shapes: (i) for pulses with rectangular shape, $f(t)= f_0 \equiv
{\rm const} \ne 0$ for $|t|<\tau_p/2$ and $f(t)=0$ otherwise, (ii)
for smooth pulses of the shape suggested by Rosen and
Zener~\cite{PhysRev.40.502}:
\begin{equation}\label{pulse}
f(t) =  \frac{\mu}{\cosh{(\pi t/\tau_p)}},
\end{equation}
where the coefficient $\mu$ is a measure of the pulse electric field
strength. The effective areas of these pulses are equal to $\Theta
= 2 f_0 \tau_p$ and $\Theta = 2 \mu \tau_p$, respectively.

For $f(t)$ taken in the form of Eq. (\ref{pulse}) one can write the solution of
Eq.~\eqref{single} following  Ref.~\cite{PhysRev.40.502}
in terms of the hypergeometric
function \begin{multline}\label{sol}
 \psi_{1/2}(t) = \psi_{1/2}(-\infty)\, \times \\
\hF\left[\frac{\Theta}{2\pi},-\frac{\Theta}{2\pi};
 \frac{1}{2}- \mathrm i y;\frac{1}{2}\tanh{\left(\frac{\pi t}{\tau_p}\right)} +\frac{1}{2}  \right],
\end{multline}
where $\Theta=2\mu\tau_p$ and the dimensionless detuning $y=
\omega'\tau_p/(2\pi)$. This leads to explicit expressions for $Q$
and $\Phi$:
\begin{eqnarray}\label{q:rosen}
 Q &=& \left|\frac{\Gamma^2 \left( \frac{1}{2} - \mathrm i y \right)}{\Gamma\left(\frac{1}{2}
 - \frac{\Theta}{2\pi} - \mathrm iy\right)\Gamma\left(\frac{1}{2} + \frac{\Theta}{2\pi}
 - \mathrm iy\right)} \right| \nonumber  \\
&=&\sqrt{1 - \frac{\sin^2(\Theta/2)}{\cosh^2{(\pi y)}}}\:,
\\
 \Phi &=& \arg{\left\{ \frac{\Gamma^2 \left( \frac{1}{2} - \mathrm i y \right)}{\Gamma\left(\frac{1}{2}
 - \frac{\Theta}{2\pi} - \mathrm iy\right)\Gamma\left(\frac{1}{2} + \frac{\Theta}{2\pi}
 - \mathrm iy\right)}\right\}}\:.
\label{eq:22}
\end{eqnarray}

In the case of a rectangular shaped pulse we obtain
\begin{eqnarray}\label{q:rect}
Q &=&\sqrt{1-\frac{\Theta^2}{x^2}\sin^2{\frac{x}{2}}}\:,
\\ \Phi &=& \pi y-\phi\:, \label{eq:26} \end{eqnarray}
where $\Theta = 2 f_0 \tau_p$, effective Rabi frequency
\begin{equation}
x = \sqrt{(2 \pi
y)^2 + \Theta^2}, 
\end{equation}
and $\sin{\phi}=(y/Qx)\sin(x/2)$. One can see from Eqs.~\eqref{eq:22} and \eqref{eq:26} that $\Phi$
changes its sign with reversal of the detuning parameter $y$. For
circularly polarized pump pulses, this means that the sign of
detuning determines the rotation direction of the electron spin
polarization and its reversal is similar to switching the
effective magnetic field from one direction to the opposite.

According to Eqs.~\eqref{pm} and \eqref{Jz} the reorientation of
electron spin polarization by short pulses with $Q=0$ or $Q=1$
becomes deterministic and can be used for controllable
manipulation of a single electron spin. The pump pulses with $Q=0$
completely erase, for any detuning, the in-plane spin components
\cite{Greilich_PRL06,Economou_PRL07} leading to the alignment of
the electron spin along the $z$ axis. On the other hand, the
pulses with $Q=1$ lead to a controllable rotation of the electron
spin in the $(x,y)$ plane and make no effect on the spin $z$
component. The rotation angle equals to $\Phi$ and is determined
by the detuning~\cite{Economou_PRL07}. It follows from
Eqs.~\eqref{q:rosen}--\eqref{eq:26} that the value $Q = 0$ can be
reached only with the pulses tuned to the resonance ($y=0$) and
having the areas $\Theta = \pi,3\pi,...$ (the so-called
$\pi$-pulses). In the case of Rosen\&Zener pulses
\cite{Economou_PRL07} the condition $Q=1$ is realized for any
detuning if $\Theta = 2 \pi, 4\pi,...$ ($2 \pi$-pulses). For the
rectangular pulses this condition can be reached only for certain
combinations of the detuning and pulse area when {the effective Rabi frequency} $x =
\sqrt{(2 \pi y)^2 + \Theta^2} = 2 \pi N$ with $N$ being an
integer.

\begin{figure}[hptb]
\includegraphics[width=\linewidth]{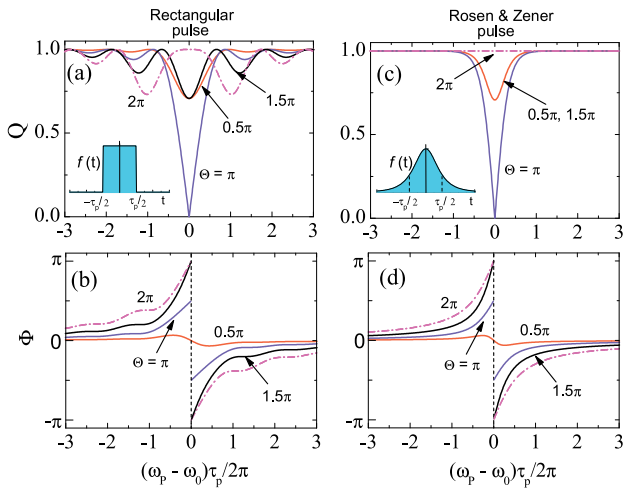}
\caption{Dependence of $Q$ and $\Phi$ on  detuning $y=(\omega_{ \mbox{}_{\rm P}}-\omega_0)\tau_p/2\pi$ for
pulses having the rectangular shape (a,b) and Rosen\&Zener shape (c,d) calculated for several pulse
areas $\Theta = \pi/2, \pi, 3\pi/2, 2\pi$. Insets demonstrate shapes of pulses. }
\label{fig:q_phi}
\end{figure}

Figure~\ref{fig:q_phi} shows the calculated dependences of $Q$ and $\Phi$ on
the detuning  for four pulse areas $\Theta$, each for
rectangular and Rosen\&Zener pulse shapes. For large detuning,
$y=|\omega_{ \mbox{}_{\rm P}}-\omega_0|\tau_p/2\pi \gg 1$, $Q$ is close to $1$,
$\Phi$ tends to $0$. Therefore, the electron spin state in a QD is
unaffected by the strongly detuned pulses. For the area $\Theta =
\pi$, the function $Q(y)$ has a sharp dip at $y=0$ reaching zero
value at this particular point. Thus, for $\pi$-pulses tuned to
the resonance $\omega_0$ the parameter $Q$ vanishes and, as stated
above, the pump pulses suppress the transverse spin components
$S_{x}^+$ and $S_{y}^+$. Deviation of $\Theta$ from $\pi$ converts
the dip into a smooth minimum.
In accordance with Eq.~\eqref{q:rosen}, the pulses with the areas
$\Theta$ and $2\pi-\Theta$ produce the same $Q$. The signs of the
phase $\Phi$ and detuning are opposite, and $\Phi$ makes, at zero
detuning, an abrupt jump from its positive maximum value to the
negative minimum value. The altitude of this jump rises along with
the pump pulse area. Although the both pulse shapes lead to
generally similar dependences of $Q$ or $\Phi$ on the detuning,
the rectangular shape pulses give rise to additional oscillations
on the curves $Q(y), \Phi(y)$ clearly seen in
Figs.~\ref{fig:q_phi}(a), \ref{fig:q_phi}(b). These oscillations are connected with
the oscillating character of the Fourier transform of a
rectangular shaped signal. They are absent for the Rosen-Zener pulse, for the general case of a smooth pulse these oscillations are much weaker than for the rectangular pulse. One can see that, for $\Theta \neq \pi$,
$Q(y)$ has several minima at $y \neq 0$. This occurs because at these detuning the {effective Rabi frequency, $x$,}  approaches to the $N\pi$ with $N=1,3, 5...$. 

\begin{figure*}[hptb]
\includegraphics[width=0.8\linewidth]{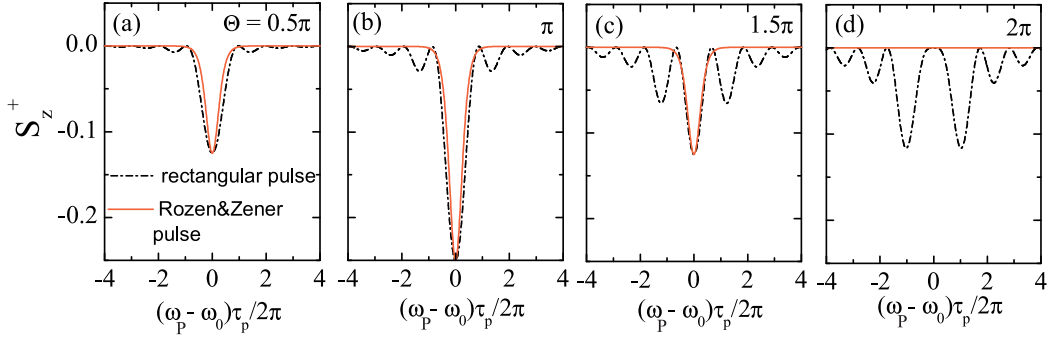}
\caption{Effect of detuning $\omega_{ \mbox{}_{\rm P}} - \omega_0$ on the $z$-component
of electron spin polarization, $S_z^+$, created by the rectangular
(dash-dot black curves) and Rosen\&Zener (solid red curves) pulses. Curves are
calculated using Eq.~(\ref{szpm}) with $S_z^- = 0$ and Eqs.
(\ref{q:rosen}, \ref{q:rect}) for the pulse areas: $\Theta= 0.5\pi$
(a), $\pi$ (b), $1.5\pi$ (c), and $2\pi$ (d). } \label{fig:sz_p}
\end{figure*}

Figure~\ref{fig:sz_p} compares the $z$-component $S_z^+$ of
electron spin polarization created by rectangular (dash-dot black
curves) and Rosen\&Zener (solid red curves) pulses of different
areas $\Theta$ in singly charged QDs with zero spin, $S_z^-=0$,
before the pulse arrival. On can see that the rectangular pulses
with $\Theta > \pi$ result in intensive oscillations of $S_z^+$ at
large-scale detuning, whereas oscillations are completely absent
for the Rosen\&Zener pulses. The oscillations are again connected
with the shape of the Fourier transform of a rectangular signal. One can see in Fig.~\ref{fig:sz_p}(d) that detuned pulses of a different shape create a completely different electron spin polarization. 

\subsection{Temporal dynamics of electron and trion spin polarization after the pulse}

The temporal dynamics of electron and trion spin polarization in a
QD after its excitation by the short pump pulse can be determined
from the kinetic equations for the electron spin polarization $\bm
S$ and  $z$-component $J_z$ of the trion spin polarization
~\cite{Shabaev_PRB03,Greilich_PRL06,Zhukov_PRB07}
\begin{eqnarray}\label{kinetic}
\dot{\bm S} + \bm S \times \bm \Omega + \frac{\bm S}{\tau_{s,e}}
&=& \frac{J_z \bm o_z}{\tau_{QD}}\:, \nonumber \\
\dot J_z + \frac{J_z}{\tau_{QD}} + \frac{J_z}{\tau_{s,h}} &=& 0\:.
\end{eqnarray}
The equations take into account (i) the precession of electron
spins in the in-plane magnetic field $\bm B$ with 
frequency $\bm \Omega$, (ii) the electron spin
relaxation, (iii) the spin relaxation of a hole in the trion, (iv)
the radiative decay of the trion, and, finally, (v)  the partial suppression of the electron spin polarization created by the pulse after the trion recombination.  {It is ignoring, however, the hole precession.} Here
$\tau_{s,e}$, $\tau_{s,h}$ and $\tau_{QD}$ are the single electron
spin relaxation time, the hole spin relaxation time and the trion
lifetime, respectively.

The detailed analysis of the spin dynamics in the coupled
electron-trion system has been carried out in
Refs.~\cite{Shabaev_PRB03,Zhukov_PRB07} (see
also~\cite{gi_08,Astakhov_SST08}). It is instructive to consider
the simplest case of these dynamics when the hole-in-trion
spin relaxation is much faster than the trion radiative lifetime,
$\tau_{s,h} \ll \tau_{QD}$, or when the electron spin precession time  $2\pi/\Omega\ll \tau_{QD}$. In the both limits the electron remaining in the QD after the trion recombination becomes completely
depolarized and its contribution to the electron spin polarization
created during the pulse is completely negligible
\cite{Greilich_PRL06,Greilich_Science06}. In this case the
precession of electron spin polarization in a transverse magnetic
field after the trion decay is described by the following set of
equations \cite{Greilich_Science06}
\begin{eqnarray}
 S_z(t) &=& [S_z^+\cos{\Omega t} + S_y^+\sin{\Omega t}]\mathrm e^{-t/\tau_{s,e}}\:,\nonumber\\
 S_y(t) &=& [S_y^+\cos{\Omega t} - S_z^+\sin{\Omega t}]\mathrm e^{-t/\tau_{s,e}}\:,\nonumber\\
S_x(t)&=&S_x^+e^{-t/\tau_{s,e}}\:, \label{prcs}
\end{eqnarray}
where time $t$ is referred to the end of excitation pulse, and the
electron spin polarization components $S_{\alpha}^+$ $(\alpha =
x,y,z)$ created by the pulse are defined by Eqs. \eqref{pm}.

\subsection{Electron spin polarization created in QDs by {an infinite} train of short pulses}
In the pump-probe Faraday and Kerr rotation experiments the sample
is usually subjected to a train of pump pulses that follow with a
certain repetition period $T_{R}$. If the time $T_{R}$ is
comparable with or smaller than the single electron spin
relaxation time in a QD, $T_R\leq\tau_{s,e}$, the electrons retain
the memory of being exposed to the previous pulses. The infinite
train of pulses creates a steady state of the electron spin
polarization  in the QDs periodically varying in time with the
same period $T_{R}$, which leads to a number of remarkable
phenomena such as the resonant spin amplification 
{\cite{Kikkawa_PRL98, gi_08}} and the mode locking of electron
spin coherence~\cite{Greilich_Science06,Greilich_PRB07}. The time
evolution between the pulses is described by Eq.~(\ref{prcs})
where $S^+_{\alpha}$ ($\alpha = x, y, z$) are the components of
spin polarization taken at the end of any pump pulse. To find
these components one should associate the polarization
(\ref{prcs}) at the moment $t = T_{R}$ with the spin polarization
at the moment before arrival of the next pulse, ${\bm S}(T_R)
\equiv {\bm S}^-$, and interconnect it with ${\bm S}^+$ according
to Eqs.~(\ref{pm}). As a result we obtain self-consistent
equations for the components $S^+_{\alpha}$. Solving them and
substituting the solution into Eqs.~(\ref{pm}) we find the
components $S^-_{\alpha}$ which can be written as
\begin{eqnarray}
S_x^- &=& K S_y^- \:, \label{modelock} \nonumber\\
S_y^- &=& \frac{ 1-Q^2}{4 \Delta} {\rm e}^{ -T_R\over\tau_{s,e} } \sin(\Omega T_R)\:, \nonumber\\
S_z^- &=& \frac{ 1-Q^2}{4 \Delta} {\rm e}^{ -T_R\over\tau_{s,e} } \times \nonumber
\\
&&\left[
Q (\cos{\Phi}-K \sin{\Phi}) {\rm e}^{ -T_R\over\tau_{s,e} } - \cos(\Omega T_R) \right]~,\nonumber\\
\end{eqnarray}
where
\begin{eqnarray}
\Delta &=& 1 - {\rm e}^{-T_R/\tau_{s,e}} \times \nonumber \\
&&\left[ \frac{1+Q^2}{2} + Q( \cos{\Phi} - K \sin{\Phi})
\right] \cos(\Omega T_R) \nonumber
\\
&+& \frac{Q (1+Q^2) }{2} {\rm e}^{-2T_R/\tau_{s,e}} ( \cos{\Phi} - K \sin{\Phi} ) \:,\nonumber
\\ K &=& \frac{Q {\rm e}^{-T_R/\tau_{s,e}} \sin{\Phi}}{1-Qe^{-T_R/\tau_{s,e}}\cos{\Phi}}\:.
\end{eqnarray}

\begin{figure}
\includegraphics[width=\linewidth]{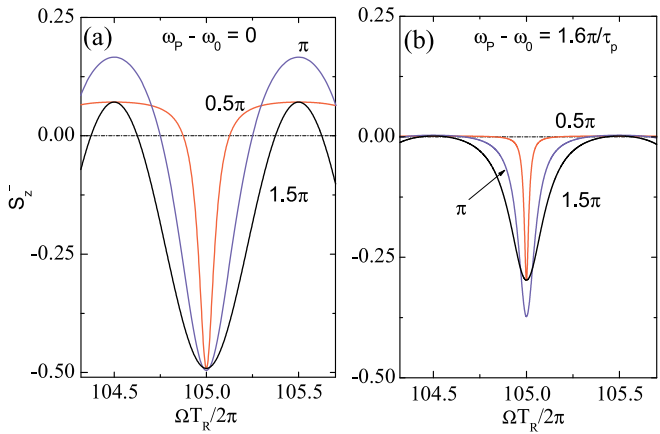}
\caption{ Electron spin polarization, $S_z^-$, created by a train of
Rozen\&Zener pump pulses with the repetition frequency $T_R=13.2$\,
ns at the moment of pulse arrival as a function of the electron spin precession
frequency $\Omega$. Dependences shown on panels (a)
and (b) are calculated, respectively, for zero detuning between the pump frequency and quantum
dot frequency, $\omega_{\mbox{}_{\rm P}}=\omega_0$, and for the detuning
$\omega_{\mbox{}_{\rm P}} - \omega_0 = 1.6\pi/\tau_p$. The chosen pulse areas are
$\Theta = 0.5\pi, \pi$ and $1.5 \pi$. For $g$-factor from
Ref.~\cite{Greilich_Science06} the value of $\Omega T_R/2\pi=105$ is reached at a
magnetic field $B=1$~T. The calculations used $\tau_{s,e}=3$~$\mu\mbox{s}$ from
Ref.~\cite{ Greilich_Science06}.}\label{fig:sz:modlock}
\end{figure}

One can check that, for a periodic train of pulses of arbitrary
intensity and shape, the electron spin polarization reaches the
highest value at the magnetic field satisfying the PSC condition
 $\Omega=2\pi
N/T_R$, where $N$ is an
integer~\cite{Kikkawa_PRL98,gi_08,Greilich_Science06,Greilich_PRB07}.
For such electrons $\cos{(\Omega T_R)} = 1$ and the
$z$-component of their spin polarization at the moment of pulse
arrival can be written as
\begin{equation}
 \label{szmax}
S_z^- = -\frac{1}{2}~\frac{1-Q^2}{2e^{T_R/\tau_{s,e}} - 1 - Q^2}\:.
\end{equation}
Equation \eqref{szmax} shows that the maximum value of $|S_z^-|$
is independent of the phase shift $\Phi$ between
$\psi_{1/2}(\infty)$ and $\psi_{1/2}(-\infty)$. For pulses with
$Q=1$ the orientation of electron spins does not occur since such
pulses rotate the in-plane spin components but do not generate the
spin coherence. Quite often  the {pulse} repetition
period used in experiments is much shorter than spin relaxation time: $T_R \ll
\tau_{s,e}$~\cite{Greilich_PRL06}. This allows to rewrite Eq.
\eqref{szmax} as
\begin{equation} \label{szmax1}
S_z^- \approx -\frac{1}{2}\frac{1}{1+2T_R/[\tau_{s,e}(1 - Q^2)]}\:.
\end{equation}
One can see that even in the case of weak excitation where $1 -
Q^2 \ll 1$ the electron spin satisfying the PSC reaches its
utmost alignment $S_z^- \approx -1/2$ if \be 2
\frac{T_R}{\tau_{s,e}} \ll 1-Q^2 = \frac{\sin^2(\Theta/2)}{
\cosh^2(\pi y)}. \label{alignment} \ee 
The latter equality is valid for Rosen\&Zener pulses. For
the large ratio $\tau_{s,e}/T_R$ even quite detuned pulses, e.g.,
with $(\omega_{\mbox{}_{\rm P}}-\omega_0)\tau_p \sim 3$, are still quite efficient
in the spin alignment. A train of weak pulses, however,
synchronizes electron spin precession only in a very narrow
frequency range around the PSC. As a result the frequency
dependence of $S_z^-(\Omega)$ created by such a train has
a periodic form with sharp minima at the frequencies satisfying
the PSC
\bea
S_z^- &\propto& 
\left[(\Omega T_R - 2\pi N)^2 + \frac{T_R^2}{\tau_{s,e}^2}\right.\nonumber\\
&+& \left.\Phi^2 + (1-Q)^2 + 2\frac{T_R}{\tau_{s,e}} (1-Q)  \right]^{-1}\:.
\eea
Here it is assumed that $1 - Q \ll 1$, $T_R \ll \tau_{s,e}$ and
$\Phi \ll 1$. The width of the minima is proportional to
$$
\frac{1}{T_R} \sqrt{ \left( \frac{T_R}{\tau_{s,e}} \right)^2
+ 2(1-Q) \frac{T_R}{\tau_{s,e}} + (1-Q)^2 + \Phi^2}\:,
$$ i.e., it is determined either by
the spin relaxation rate $\tau_{s,e}^{-1}$ or by the effective
pump area $1-Q$ and phase $\Phi$. Figure \ref{fig:sz:modlock}
shows just one period of this dependence.

The modulation of the electron spin polarization
$S_z^-(\Omega)$ becomes weaker with the increasing
detuning. For example, assuming $\Phi = \pm \pi/2$ and using the
condition $T_R\ll \tau_{s,e}$ one can derive for arbitrary $Q$ \be
 S_z^-(\Omega) = -\frac{1}{2}~ \frac{Q^2+\cos{\Omega T_{R}}}{Q^2+2-\cos{\Omega T_{R}}}\:,
\ee which is a much smoother function of $\Omega$.

For the $\pi$-pulse excitation ($Q \to 0$) which can be realized
only in the absence of detuning ($\Phi=0$) we arrive at
\cite{Greilich_Science06} \be S_z^- =
-\frac{1}{2}~\frac{\cos{\Omega T_{R}}}{2-\cos{\Omega T_{R}} }\:. \ee The polarization $S_z^-$ reaches its minimum
value $-1/2$ when the electron spin precession frequency satisfies
the PSC condition.

For a given pulse shape the phase $\Phi$ and the factor $Q$ are
interconnected and, in general, a change of the pulse area results
in changes of both $Q$ and $\Phi$. Figure~\ref{fig:sz:modlock}
shows the $z$ component of electron spin polarization, $S_z^-$, at
the moment of pulse arrival calculated as a function of the Larmor
frequency for different pulse areas under resonant excitation and
for nonzero detuning. One can see from
Fig.~\ref{fig:sz:modlock}(a) that for zero detuning ($\Phi=0$)
and small pulse areas, $S_z^-$ exhibits a sharp minimum as a
function of Larmor precession frequency in agreement with the
above analytical considerations. The increase of the pulse area
transforms this minimum to the cosine-like curve.

Figure~\ref{fig:sz:modlock}(b) shows $S_z^-(\Omega )$
created by an appreciably detuned pulse train, $\omega_{\mbox{}_{\rm P}} -
\omega_{0} = 1.6\pi/\tau_p$. In agreement with
Fig.~\ref{fig:q_phi} the effective pump power in this case is
small, i.e., $Q$ is close to $1$. An increase of the pump area
from $\pi/2$ to $\pi$ leads to a small decrease of $Q$ and a
nonzero value of $\Phi$. As a result the minimum at the spin
precession frequency $\Omega $ satisfying the PSC becomes
deeper and wider. The further increase of the pump area results in
additional widening of the minimum due to the increase of $\Phi$
but the its depth becomes smaller since $Q$ starts to increase.

\section{Probing spin dynamics in quantum dots}\label{sec:probe}

The detection of the QD spin polarization in pump-probe Faraday
and Kerr rotation experiments is carried out by a linearly
polarized probe pulse. The electric field of the probe pulse shown in Fig.~1 oscillates along the $x$ axis and similarly to Eq.
\eqref{elfield} it can be written as
\begin{equation}
 \label{probe:x}
\bm E^{\rm pr}(\bm r,t) = E_x^{\rm pr}(\bm r,t) \frac{\bm o_+ + \bm o_-}{\sqrt{2}} + {\rm c.c.}
\end{equation}
Here we assume that $E_x^{\rm pr}(\bm r, t) \propto \mathrm
e^{-\mathrm i \omega_{\rm pr} t}$, where $\omega_{\rm pr}$ is the
carrying frequency of the probe light. In order to calculate
Faraday and Kerr rotation angles of polarization plane of the
probe pulse as well as the ellipticity we first find the QD
polarization induced by the probe field and then calculate the
secondary electric field induced by the QD array.

\subsection{Probe-induced polarization of QDs}

Before the probe pulse arrival the electronic state of a QD is
described by the wavefunction \eqref{wave}. We consider the
general case where the QD is characterized by the population of the electron $n_e
= |\psi_{1/2}|^2 + |\psi_{-1/2}|^2$  and trion $n_t = |\psi_{3/2}|^2 + |\psi_{-3/2}|^2$
states and the spin polarization of these states $S_z =
\left(|\psi_{1/2}|^2-|\psi_{-1/2}|^2\right)/2 \ne 0$ and $J_z =
\left(|\psi_{3/2}|^2-|\psi_{-3/2}|^2\right)/2 \ne 0$,
respectively. Solving Eqs.~\eqref{system:1}, \eqref{system:2} in
the lowest order in $\bm E^{\rm pr}$ we find the probe-induced
corrections to the electron and trion components of the wave
function
\begin{eqnarray}  \label{sol1}
\delta \psi_{\pm 3/2} &=& \psi_{\pm 1/2} \int_{-\infty}^t \frac{V(t')}{\mathrm i\hbar}
\mathrm e^{-\mathrm i \omega_0 (t -t')}  \mathrm dt'\:,\nonumber \\
\delta\psi_{\pm 1/2} &=& \psi_{\pm 3/2}  \int_{-\infty}^t
\frac{V^*(t')}{\mathrm i\hbar} \mathrm e^{\mathrm i \omega_0 (t -t')} \mathrm dt'\:,
\end{eqnarray}
where
\begin{equation}
 V(t) = -\frac{1}{\sqrt{2}} \int \mathsf d(\bm r) E_x^{\rm pr}(\bm r, t) \mathrm d^3 r\:.
\end{equation}

The electron-trion superposition excited in a QD by the light
pulses creates a local polarization, whose magnitude depends on
the components of the wave function described in Eq.~\eqref{wave}.
According to the selection rules, the circular components of the
QD dielectric polarization can be written as
\bea
P_{\sigma^+}(\bm r) &=& \mathsf d^*(\bm r) (\psi_{3/2} + \delta \psi_{3/2})
 (\psi_{1/2} +\delta \psi_{1/2})^* + {\rm c.c.}\:,\nonumber\\
P_{\sigma^-}(\bm r) &=& \mathsf d^*(\bm r) (\psi_{-3/2} + \delta \psi_{-3/2})
 (\psi_{-1/2} + \delta \psi_{-1/2})^* \nonumber\\
&+& {\rm c.c.}\:, \label{eq:35}
\end{eqnarray}
where the effective transition dipole is defined by Eq. \eqref{dpm}. In Eqs.~\eqref{eq:35} the zero-order contributions, which are
proportional to $\psi_{3/2}\psi_{1/2}^*$ and
$\psi_{-3/2}\psi_{-1/2}^*$, determine the QD emission due to the
presence of photoexcited trions. They make no contribution to the
measured pump-probe rotation signal and will not be considered further. The
other contributions in Eqs.~\eqref{eq:35} are induced by the probe
pulse. The Faraday and Kerr rotation of the probe light
polarization as well as its ellipticity are determined only by the
terms linear in $\delta \psi_{\pm 1/2}$ and $\delta \psi_{\pm 3/2}$.
Combining Eqs.~\eqref{eq:35} we can write the
linearly polarized components of the QD polarization induced by
the probe pulse as follows
\begin{widetext}
\begin{eqnarray}
&&\delta  P_{x}^{QD}(\bm r,t) = - \frac{n_e - n_{tr}}{2 \mathrm i \hbar} \mathsf d^*(\bm r) \int
\mathrm d^3 r' \int\limits_{-\infty}^t  \mathrm dt'\mathrm e^{\mathrm i \omega_0
(t' -t)} \mathsf d(\bm r') E_x^{\rm pr}(\bm r',t') + {\rm c.c}\:,\label{px1}\\
&&\delta  P_{y}^{QD}(\bm r,t) = -\frac{S_z - J_z}{\hbar}  \mathsf d^*(\bm r) \int \mathrm d^3r'
\int\limits_{-\infty}^t \mathrm dt'\mathrm e^{\mathrm i \omega_0 (t' -t)} \mathsf d(\bm r')
E_x^{\rm pr}(\bm r',t')  + {\rm c.c.}\:. \nonumber
%\label{py1}
\end{eqnarray}
The light wavelength is usually much larger than the size of
self-organized QDs. This allows one to extract the probe electric
field $E_x^{\rm pr}(\bm r',t')$ from the integral and present the
QD polarization in the approximate $\delta$-function-like form
\[
\delta  P_{\alpha}^{QD}(\bm r,\bm R_j, t) = \delta(\bm r - \bm R_j) \Pi_{\alpha}(\bm R_j,t)
\hspace{5 mm} (\alpha= x,y) \:,
\]
where $\bm R_j$ is the position of $j$-th QD. The resulting probe-field induced polarization of a
single QD can be expressed as
\bea \Pi_x(\bm R_j,t) &=& - \frac{n_e -
n_{tr}}{2\mathrm i \hbar} |\mathcal D|^2 \int_{-\infty}^t
\mathrm e^{\mathrm i \omega_0 (t' -t)} E_x^{\rm pr}(\bm R_j,t') \mathrm d t' + {\rm c.c.}\:,\label{px2}\\
\Pi_y(\bm R_j,t) &=& - \frac{S_z - J_z}{\hbar} |\mathcal
D|^2\int_{-\infty}^t \mathrm e^{\mathrm i
\omega_0 (t' -t)} E_x^{\rm pr}(\bm R_j,t') \mathrm d t' + {\rm
c.c.}\:,\label{py2} 
\eea  
via the integral QD transition dipole $\mathcal D =\int \mathrm d^3 r \, \mathsf d(\bm r)$
related to the two-particle wave function ${\mathsf F}({\bm r}, {\bm r})$:
$$
|\mathcal D|^2 = \frac{e^2 |p_{cv}|^2}{\omega_0^2 m_0^2} \left|\int {\mathsf F}(\bm
r, {\bm r}) \mathrm d^3 r \right|^2\:.
$$
\end{widetext}

As one can see from Eqs.~{\eqref{px1}} the probe-induced QD
polarization $\delta  {\bm P}^{QD}$ consists of two components.
The first, $\delta P_{x}^{QD}$, is parallel to the probe
polarization plane and its magnitude is proportional to the
difference of electron and trion occupation numbers. The second
component, $\delta P_y^{QD}$, is orthogonal to the probe
polarization plane and its magnitude is proportional to the
difference of the electron and trion spin polarizations. The
latter polarization component is responsible for the probe-pulse
polarization plane rotation, i.e., spin Faraday and Kerr effects,
and for circular dichroism (ellipticity of the transmitted or
reflected probe beam). Note that the appearance of the $\delta
P_y^{QD}$ component is not a direct consequence of the spin-orbit
interaction: this component is not relativistically small as
compared with $\delta P_x^{QD}$.

\subsection{Circular birefringence and dichroism induced by photoexcited QDs}

Once the probe-induced dielectric polarization of the QD is known,
it is possible to calculate an electric field induced by the QD
ensemble and, therefore, find the probe polarization plane
rotation and ellipticity. First we consider an experimental
situation where pumping and probing are carried out on a planar
array of QDs. Then we generalize the results to a stack of such QD
planes and a bulk array of QDs.

Let us consider a layer of self-organized QDs forming the plane
$z=0$. The total electric field $\bm E$ \cite{Onecomponent} in the system can be
represented as a sum of the incident electric field $\bm E_0^{\rm
pr}(t) e^{\mathrm i qz}$ and the electric field
induced by the QD dielectric polarization $\delta  {\bm P}^{QD}$.
The field $\bm E$ satisfies the electromagnetic wave equations
\begin{eqnarray}  \label{max1}
\Delta \bm E(\bm r,t) - \grad\divv \bm E(\bm r,t) &=& -
\left(\frac{\omega_{\rm pr}}{c} \right)^2 \bm D\:,\\
\divv \bm D &=& 0\:, \label{Disp}
\end{eqnarray}
with the material equation
\begin{equation} \label{divD}
\bm D(\bm r,t) = \varepsilon_b \bm E(\bm r,t) + 4\pi \bm
P_{\rm tot}(\bm r,t)\:.
\end{equation}
Here $\varepsilon_b$ is the dielectric constant of the cap layer
assumed to coincide with the background dielectric constant of the QDs; $\bm P_{\rm tot}(\bm r, t)=\sum_j \delta\bm P^{QD}(\bm r,\bm R_j,t)$ is the sum of the probe-induced
polarizations over all QDs; $\bm D$ is the displacement field;
$\omega_{\rm pr}$ is the carrying frequency; and $c$ is the speed
of light in vacuum. Although the pumping and probing of QDs is
performed by short pulses, their duration $\tau_p$ is assumed to
exceed by far the period of electro-magnetic field oscillations $2
\pi/\omega_{\rm pr}$. Therefore, the solutions of
Eqs.~(\ref{max1})-(\ref{divD}) are quasi-monochromatic waves with
slowly varying amplitudes.

It follows from Eq.~\eqref{Disp} that $\divv \bm E =
-(4\pi/\varepsilon_b)\divv \bm P_{\rm tot}(\bm r)$ which allows us
to rewrite Eq.~(\ref{max1}) in the form
\begin{multline}
 \label{max:single}
\Delta \bm E(\bm r,t) + q^2 \bm E(\bm r,t) = \\ - 4 \pi \left(\frac{\omega_{\rm pr}}{c} \right)^2
(1+ q^{-2} \grad\divv)\bm P_{\rm tot}(\bm r,t)\:,
\end{multline}
where {$q = \omega_{\rm pr}\sqrt{\varepsilon_b}/c$}. By
introducing the Green's function for the three-dimensional space
\begin{equation}
 \label{greens:3d}
G(\bm r) = \frac{\exp{(\mathrm i q r)}}{4\pi r}\:,
\end{equation}
Eq.~(\ref{max:single}) can further be transformed into an integral
equation
\begin{multline}
 \label{maxwell:gen}
\bm E(\bm r,t) = \bm E_0^{\rm pr}(t) e^{\mathrm i qz} + \\ 4\pi \left(\frac{\omega_{\rm pr}}{c} \right)^2 \int
\mathrm d^3 r' G(\bm r - \bm r') (1 + q^{-2} \grad\divv)\bm P_{\rm tot}(\bm r',t)\:.
\end{multline}

\begin{figure}[hptb]
\includegraphics[width=\linewidth]{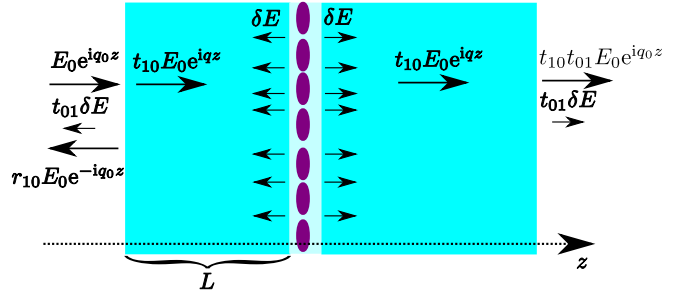}
\caption{ Schematic image of the light propagation in a QD structure,
which contains a QD layer overgrown by a cap layer of the thickness $L$.
Long arrows show the transmission and reflection of the incident probe light (with the electric-field amplitude
$E_0$) on the external surface while short arrows illustrate the creation of the secondary field, $\delta E$,
due to scattering of the transmitted light by QDs. $r_{10}$, $t_{10}$ and ${t}_{01}$ are the
corresponding reflection and transmission coefficients.
 }\label{fig:array}
\end{figure}

A plane wave of the probe electromagnetic field propagating along
the $z$ direction creates a dielectric polarization in QDs
randomly distributed in the plane $z=0$. Assuming the interdot
distances to be smaller than the light wavelength one can neglect
the $q^{-2}\grad\divv$ term in the integral of Eq.
\eqref{maxwell:gen} and average $\bm P_{\rm tot}(\bm r)$ over the
distribution of QDs. As a result we can replace $\bm \Pi(\bm
R_j,t)$ by the coordinate independent vector $\bm \Pi(t)$. This
allows us to rewrite Eq.~\eqref{maxwell:gen} in the following form
\begin{multline}
 \label{eq:48}
\bm E(\bm r,t) = \bm E_0^{\rm pr}(t) e^{\mathrm i qz} - \\4\pi \left(\frac{\omega_{\rm pr}}{c}
\right)^2 \bm \Pi(t)\sum_j \int \mathrm d^2 \rho'\mathrm dz' G(\bm r - \bm r')
\delta(\bm \rho_j-\bm\rho')\delta(z')\:,
\end{multline}
where $\bm \rho_j$ are the QD positions in the two-dimensional
layer. Assuming that QDs in the array are identical and randomly
distributed we replace the sum in Eq.~\eqref{eq:48} by the
integral and arrive at
\begin{equation}
 \label{maxwell:qd}
\bm E(\bm r,t) = \bm E_0^{\rm pr}(t) e^{\mathrm i qz} - 4\pi \left(\frac{\omega_{\rm pr}}{c}
\right)^2 {\frac{\mathrm i }{2q}}{\rm e}^{\mathrm i q|z|} N_{QD}^{2{D}}  \bm \Pi (t)\:,
\end{equation}
where $N_{QD}^{2{D}}$ is the two-dimensional density of QDs. Deriving Eq. \eqref{maxwell:qd} we used the following property of
two-dimensional integral $$\int 2\pi\rho\mathrm d\rho
\frac{\exp(\mathrm
iq\sqrt{z^2+\rho^2})}{4\pi\sqrt{z^2+\rho^2}}=\frac{\mathrm
i}{2q}\exp(\mathrm iq|z|)\:,$$ which can be proven by adding a
small positive imaginary part to $q$ and setting it to $+ 0$.

It is convenient to represent the electric field in a QD sample as
\begin{equation}
 \label{maxwell:qd2}
\bm E(\bm r,t) = \bm E_0^{\rm pr}(t) e^{\mathrm i q z} +
\delta \bm E(t) \mathrm e^{\mathrm i q |z|}\:,
\end{equation}
where the first term is just the incident probe field and the
second term describes the secondary field induced by QDs:
\begin{equation}
 \label{deltaE}
\delta \bm E(t) = - 4 \pi \left(\frac{\omega_{\rm pr}}{c}
\right)^2 \frac{\mathrm i}{2q}N_{QD}^{2{D}} \bm \Pi (t)\:.
\end{equation}
This equation allows one to find magnitudes of the Faraday
rotation signals and ellipticity.

If the sample contains $M$ layers of QDs and the stack thickness
$d$ is smaller than the light wavelength then the second term in
Eq.~\eqref{maxwell:qd2} should merely be multiplied by $M$. In a
more conventional description of three-dimensional (3D) ensemble
of QDs, the factor $M N_{QD}^{2{D}}$ can be rewritten as
$N_{QD}^{3{D}}$, where $N_{QD}^{3{D}}$ is the 3D concentration of
QDs.

\begin{widetext}
The straightforward calculation shows that the Faraday rotation
signal defined by Eq.~\eqref{eq:1} can be presented as
\begin{equation}
 \label{farad:gen}
\mathcal F  =-2\int_{-\infty}^\infty \Re{[E_{0,x}^{\rm pr*}(t)\delta E_y(t)]}\mathrm dt \,,
\end{equation}
where only a contribution linear in $\delta \bm E$ is taken into
account. Substituting $\delta E_y$ from Eqs.~\eqref{py2} and
\eqref{deltaE} we arrive at
\begin{equation}
 \label{farad:fin}
\mathcal F = \frac{3\pi}{q^2\tau_{QD}}  N_{QD}^{2{D}} (S_z - J_z)
\Im\left\{\int_{-\infty}^\infty \mathrm d t \int_{-\infty}^t \mathrm dt'
\mathrm e^{\mathrm i \omega_0(t'-t)} E_{0,x}^{{\rm pr}*}(t) E_{0,x}^{\rm pr}(t')\right\}\:,
\end{equation}
where $\tau_{QD}$ is the radiative lifetime of an electron-hole
pair confined in a QD:
\begin{equation}
 \frac{1}{\tau_{QD}} = \frac{4}{3}\frac{q^3}{\varepsilon_b \hbar} |\mathcal D|^2\:.
\end{equation}

Similarly, the calculation of the ellipticity defined by Eq.
\eqref{eq:3} results in
\begin{equation}
 \label{ellipt:gen1}
\mathcal E  =  -2\int_{-\infty}^\infty \Im{[E_{0,x}^{\rm pr*}(t)\delta E_y(t)]}\mathrm dt\:.
\end{equation}
Substituting $\delta E_y$ from Eqs.~\eqref{py2} and \eqref{deltaE}
we obtain
\begin{equation}
 \label{ellipt:fin}
\mathcal E = \frac{3\pi}{q^2\tau_{QD}}  N_{QD}^{2{D}} (S_z - J_z)
\Re\left\{\int_{-\infty}^\infty \mathrm d t \int_{-\infty}^t \mathrm dt'
\mathrm e^{\mathrm i \omega_0(t'-t)} E_{0,x}^{{\rm pr}*}(t) E_{0,x}^{\rm pr}(t')\right\}\:.
\end{equation}
\end{widetext}

In samples with a cap layer, see Fig.~\ref{fig:array}, the Faraday
rotation and ellipticity signals acquire an extra factor
$t_{10}t_{01}$ in Eq.~\eqref{farad:fin}, where $t_{01}$ and
$t_{10}$ are the transmission coefficients through the interface from
the cap layer to vacuum and vice versa, respectively.

The cap layer strongly influences the Kerr effect, i.e., the
polarization plane rotation in the reflection geometry. This
happens because its magnitude is determined by the interference
between the probe beam reflected from the cap layer and the
secondary wave induced by the QDs, see Fig.~\ref{fig:array}. The
phase difference of the reflected and secondary waves is
determined by the cap layer thickness $L$ leading to the following
expression for the Kerr rotation magnitude
\begin{equation}
 \label{kerr:fin}
\mathcal K = r_{01}t_{01}t_{10} [\cos{(2qL)} \mathcal F + \sin{(2qL)} \mathcal E]\:,
\end{equation}
where $\mathcal F$ and $\mathcal E$ are given by
Eqs.~\eqref{farad:fin} and \eqref{ellipt:fin}, respectively, and
$r_{01}$ is the reflection coefficient from the vacuum -- cap
layer interface. It is seen that the Kerr effect measures, in
general, a superposition of the Faraday rotation and ellipticity
signals.

Equations~\eqref{farad:fin} and \eqref{ellipt:fin}
demonstrate that Faraday, ellipticity, and therefore Kerr signals (see Eq.\eqref{kerr:fin}) are
proportional to the difference of electron and trion spin
polarization in QDs: $S_z- J_z$. The magnitudes of the effects are
proportional to the QD density and increase with a
decrease of {the radiative lifetime} $\tau_{QD}$ {due to an increase of the transition dipole moment}.

In order to analyze the dependence of the Faraday and ellipticity
signals on the detuning between probe frequency, $\omega_{\rm
pr}$, and trion resonance frequency, $\omega_0$, we represent
probe field as $E_0^{\rm pr}(t) = E^{(0)} s(t) e^{-\mathrm i
\omega_{\rm pr} t}$, where $s(t)$ is the envelope function. It can
be seen {from Eqs.~\eqref{farad:fin} and \eqref{ellipt:fin}} that
\begin{equation}
 \label{freq}
\mathcal F \propto \Im{G(\omega_{\rm pr} - \omega_0)}\:,
\quad \mathcal E \propto \Re{G(\omega_{\rm pr} - \omega_0)}\:,
\end{equation}
where
\begin{equation}
 G(\Lambda) = \int_{-\infty}^{\infty} \mathrm dt \int_{-\infty}^t \mathrm dt' s(t) s(t')
 \mathrm e^{\mathrm i \Lambda(t-t')}, \end{equation}
with $\Lambda= \omega_{\rm pr} - \omega_0$.
It can be recast as a half axis Fourier transform of the probe
autocorrelation function
\[
 G(\Lambda) = \int_{0}^{+\infty} \mathrm dt\, \mathrm e^{\mathrm i \Lambda t} \int_{-\infty}^{\infty}
 \mathrm dt' s(t') s(t+t')
\]
and calculated for particular pulse shapes as follows
\begin{widetext}
\begin{equation}\label{G}
G(\Lambda) = \begin{cases} \cfrac{1}{\Lambda^2}(\mathrm i\Lambda\tau_p + 1 - {\rm e}^{\mathrm i \Lambda \tau_p}) &
{\rm for}~s(t) = 1 \hspace{2 mm} \mbox{if} \hspace{2 mm} -\tau_p/2 \le t \le \tau_p/2
\hspace{2 mm} \mbox{and} \hspace{2 mm} s(t)=0 \hspace{2 mm} \mbox{otherwise,} \\
\cfrac{\tau_p^2}{\pi^2}\zeta{\left(2,\frac{1}{2}-\frac{\mathrm i
\Lambda\tau_p}{2\pi}\right)} &{\rm for}~ s(t) = \cosh^{-1}(\pi t/\tau_p) \:, \\
\tau_p^2\cfrac{2+\mathrm i \Lambda\tau_p (3+\Lambda^2\tau_p^2)}{(1+\Lambda^2\tau_p^2)^2} &
{\rm for}~s(t) = \mathrm e^{-|t|/\tau_p} \:,\\
\end{cases}
\end{equation}
where $\zeta(a,b)$ is the generalized Riemann $\zeta$-function
defined as $\zeta(a,b) = \sum_{k=0}^\infty (k+b)^{-a}$.
\end{widetext}

\begin{figure}[hptb]
\includegraphics[width=\linewidth]{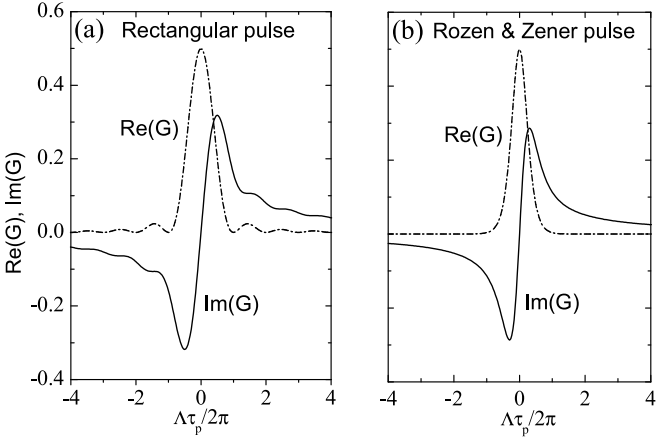}
\caption{The dependence of real and imaginary parts of $G$ on detuning,
$\Lambda=\omega_{\rm pr}-\omega_0$. Dash-dot and solid curves are calculated, respectively,
for (a) the rectangular and (b) Rosen\&Zener pulses.}
\label{fig:probe}
\end{figure}

Figure~\ref{fig:probe} shows the real and imaginary parts of $G$ calculated for the pulses of rectangular shape (panel (a)) and  Rosen\&Zener shape (panel (b)), respectively. For the  both pulse shape the dependences $G(\Lambda)$  look qualitatively  very similar. One can see that the Faraday and ellipticity
signals are, respectively, odd and even functions of the detuning.
Therefore, the ellipticity reaches its maximum sensitivity for
zero detuning, $\Lambda=0$, whereas the Faraday signal is most
sensitive for the detuned probe pulses with $|\Lambda| \tau_p \approx
1$.  The Kerr signal dependence on detuning has, in general, an
asymmetric profile due to a combined contribution from $\mathcal
F$ and $\mathcal E$ to its magnitude, see Eq.~\eqref{kerr:fin}. As
shown below, this different frequency sensitivity leads
to the different time-resolved Faraday, Kerr and ellipticity
signals in a QD ensemble with inhomogeneously broadened resonant
transition energies.

\subsection{Effective media approximation}\label{subsec:eff}

At the end of this section we establish the link between the microscopic approach developed above and semi-phenomenological effective medium approximation which is a standard tool for the description of Faraday, Kerr and ellipticity effects in bulk systems. We demonstrate below that the effective medium approximation can describe the Faraday and Kerr effects in a 3D ensemble of QDs provided that the QD density satisfies certain conditions.

The irradiation of bulk
homogeneous semiconductors with circularly polarized light creates
non-equilibrium population of electrons and holes as well as
non-equilibrium orientation of their spins and, thus, a
nonequilibrium macroscopic magnetization. Hence, the semiconductor
after the absorption of a circularly polarized pump pulse changes
its dielectric and magnetic properties. These modifications can be
tested  by the linearly polarized probe light: the polarization
plane rotates after its transmission through or reflection from
the photoexcited medium leading to the Faraday and Kerr effects,
respectively. The effects are phenomenologically described by the Fourier component of
the displacement field, $\bm D(\omega)$, which is connected with the
Fourier component of the electric field of the probe light, $\bm E(\omega)$, by
\begin{equation} \label{gyrotropy}
\bm D = \varepsilon_b \bm E + \delta\hat{\bm \varepsilon}
\bm E+ \mathrm i [\bm E \times \bm g]\:.
\end{equation}
Here $\varepsilon_b$ is the background dielectric constant,
$\delta\hat{\bm \varepsilon}$ is the spin-independent modification
of the dielectric tensor $\hat{\bm \varepsilon}$ due to the
filling of the conduction- and valence-band edge states by the
photoexcited carriers, and $\bm g$ is the gyration vector pointing
in the direction determined by the spin orientation ${\bm S}$ of
photoexcited carriers and the point-group symmetry of the system.
In bulk cubic semiconductors $\bm g \propto \bm S$ and the tensor
$\delta \hat{\bm \varepsilon}$ reduces to a scalar
$\delta\varepsilon$.

The same description can be used for the 3D ensemble of QDs if
their concentration $N_{QD}^{3{D}}$ satisfies two conditions. Firstly, $N_{QD}^{3{D}}$ should
be sufficiently small so that the QDs may be considered as
independent dipoles. Secondly, $N_{QD}^{3{D}}$ should be
sufficiently large to have the typical distances between QDs
smaller than the light wavelength. The satisfaction of these
conditions allows one to neglect the non-locality of the QD
response and represent the displacement field $\bm D$ in the QD
sample as \be \bm D(\bm r,t) = \varepsilon_b \bm E(\bm r,t)+4 \pi
\delta\bm  P(\bm r,t)\:, \ee where the optically induced
dielectric polarization is related to the electric field by
\begin{equation}
\delta\bm P(\bm r,t) = \int_{-\infty}^t \hat \varkappa(t-t') \bm E(\bm r, t') \mathrm d t'\:.
\end{equation}
Using Eqs.~\eqref{px2}, \eqref{py2} we can present nonzero
components of the tensor $\hat \varkappa(\tau)$ as
\begin{eqnarray}
 \varkappa_{xx}(\tau) = \varkappa_{yy}(\tau) =
 -N_{QD}^{3{D}}\frac{n_e - n_{tr}}{\mathrm 2i \hbar}|\mathcal D|^2
 \mathrm e^{- \mathrm i \omega_0\tau}\:, \label{fdiag} \\
 \varkappa_{yx}(\tau) = -\varkappa_{yx}(\tau)
 = -N_{QD}^{3{D}}\frac{S_z - J_z}{\hbar}|\mathcal D|^2 \mathrm
 e^{-\mathrm i \omega_0\tau}\:. \label{fnondiag}
\end{eqnarray}
It follows then that the QD contribution to the
frequency-dependent dielectric permittivity tensor can be written
as~\cite{ll8_eng} 
\bea \varepsilon_{xx} (\omega) &=&
\varepsilon_{yy} (\omega) \nonumber\\
&=& \varepsilon_b +
  (n_e - n_{tr}) \frac{2\pi N_{QD}^{3{D}} |\mathcal D|^2}{\hbar (\omega_0 -
\omega - {\rm i} 0)}\:,\nonumber\\
   \varepsilon_{yx} (\omega)  &=&  - \varepsilon_{xy} (\omega) = (S_z - J_z)
\frac{4 \pi \mathrm i N_{QD}^{3{D}}|\mathcal D|^2}{\hbar (\omega_0 -
\omega - {\rm i} 0)}\:. \eea 
Comparison with Eq. \eqref{gyrotropy}
shows that the gyration vector $\bm g$ in the photoexcited QD
medium has only one nonzero component, \be g_z = (J_z - S_z)
\frac{4\pi N_{QD}^{3{D}}|\mathcal D|^2}{\hbar (\omega_0 - \omega -
{\rm i} 0)}\:, \ee 
which is proportional to the difference of spin
densities of electrons and trions in the system and has a
resonance at the trion excitation frequency. On the other hand,
the modification of diagonal components of the dielectric tensor  $\varepsilon_{xx} =
\varepsilon_{yy}$ is proportional to the difference in population
of the electron and trion levels irrespective to the their spin
orientation.

\section{Time-dependent traces of pump-probe Kerr and Faraday rotation and ellipticity signals}\label{sec:dyn}

In this section we apply the derived general expressions to
calculate the typical time-dependent  traces of two color pump-probe Faraday and
Kerr rotation (FR and KR) signals as well as the ellipticity
created by short pulses of the resonant light and by a train of
such pulses in an ensemble of singly charged QDs. The real QD structures 
possess two important properties that affect strongly the time dependent traces, but have not been considered in
the previous sections. They are (i) inhomogeneity of a QD ensemble
expressed in dispersion of the QD resonant transition energies and
electron $g$-factors, and (ii) dispersion of electron precession frequencies connected with fluctuations of the nuclear contribution to
these frequency.  Here we conduct calculations for
the QD ensemble assuming {that the broadening of trion resonance
frequency, dispersion of  electron $g$-factors and fluctuations of the nuclear contributions to the electron spin precession frequency are  similar} to those in the samples
studied in a series of works
\cite{Greilich_PRL06,Greilich_Science06,Greilich_PRB07,Greilich_Science07}.
Those samples contained 20 layers of InGaAs QDs self-organized
during the molecular-beam epitaxy growth.
 
In the following calculations we neglect the scatter in the QD oscillator transition strengths
and the nuclear induced frequency focusing effect~\cite{Greilich_Science07}.
For illustrative purposes below we show the FR and ellipticity
signals created by the electron spin polarization only. We neglect
the trion $J_z$-dependent contribution to these signals in Eqs.
\eqref{farad:fin} and \eqref{ellipt:fin} which affects the time
dependence traces only during the trion recombination time
$\tau_{QD}\approx 400$\,ps \cite{Greilich_PRL06}. As a result the
calculated dependences can be directly compared with experimental data only
for times longer than $\tau_{QD}$.

\subsection{Modeling of inhomogeneities in a QD ensemble} 

\begin{figure}[hptb]
\includegraphics[width=0.6\linewidth]{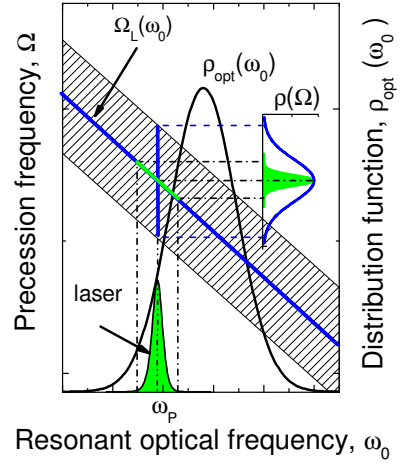}
\caption{Schematic illustration of the resonant transition
energies distribution $\rho_{\rm opt}(\omega_0)$ (black solid line) and the part of this distribution excited by short resonant pulse (green/gray profile). The range of Larmor frequencies  created by the short pulse is shown by the crosshatched region around the linear dependence
${\Omega}_{{\rm L}}(\omega_0)$.  Inset shows the distribution of the electron spin precession frequencies $\rho(\Omega)$  with (blue/solid line) and without (filled green/gray area) shape and composition contribution to the $g$-factor dispersion and nuclear fluctuations. }
\label{fig:gfactor}
\end{figure}

To model time dependences of the FR and ellipticity signals
generated by the resonant pump pulses of circularly polarized
light we assume that the distribution of the resonant transition
energies, $\rho_{\rm opt}(\omega_0)$, in the QD ensemble has the Gaussian
form:
 \begin{equation} \label{rho:omega0}
\rho_{\rm opt}(\omega_0)=\exp{ \left[ - \frac{\hbar^2(\omega_0 - \bar{\omega}_0)^2}{2 (\Delta E)^2}\right]}\:,
\end{equation}
where $\bar{\omega}_0$ is the average trion transition frequency and
$\Delta E$ is the half-width of this distribution. The distribution is shown in Fig.~\ref{fig:gfactor}.
In the
calculations we used $\hbar\bar{\omega}_0 = 1.4$~eV and
$\Delta E = 6.5$~meV from Ref.~\cite{Greilich_Science06}. Only a small part of this distribution is excited by the pump
pulse with $\tau_p=1.5$~ps. This part is proportional to the pulse spectral width  $\sim \hbar/\tau_p=1.75$~meV and is centered
at the pump carrier frequency $\omega_{\mbox{}_{\rm P}}$. The photoexcited part of the QD distribution is shown in Fig.~\ref{fig:gfactor} by 
filled Gaussian at low-energy part of $\rho_{\rm opt}(\omega_0)$.

The dispersion of the electron spin precession frequency in a QD
ensemble is determined by the dispersion of electron $g$-factors
and fluctuations of the nuclear contribution to the precession.
The electron $g$-factor, $g_e$, depends generally on the effective
energy gap of the QD, i.e. on the optical transition frequency,
as well as on the QD shape and composition~{\cite{ivchenko05a}}. The first effect gives
rise to a correlation between the average value of $g$-factor and
the trion resonance frequency and can be approximated by a linear
function
\begin{equation} \label{g:omega0}
g_e(\omega_0) = A \hbar\omega_0 + C\:,
\end{equation}
where $A$ and $C$ are fitting parameters. This results in the dependence of Larmor precession frequency,
${\bm \Omega}_{\rm L}(\omega_0) = \mu_B g_e (\omega_0) {\bm B}/\hbar$, on a trion optical 
resonance frequency, which is shown in Fig.~\ref{fig:gfactor} by thick inclined line. 
The spread of  ${\Omega}_{\rm L}$ connected with the $g$-factor dependence on an excitation frequency
is controlled by the pump pulse width $\hbar/\tau_p$. This spread is marked by  a green/gray  segment on a linear dependence
of ${\Omega}_{\rm L}(\omega_0)$ in Fig.~\ref{fig:gfactor}. The distribution of electron spin precession frequencies, $\rho(\Omega_{\rm L})$ created by this effect is shown  in the inset of  Fig.~\ref{fig:gfactor} by filled Gaussian. 
 We use in
our calculations $A = - 1.75$~$\mu$eV$^{-1}$ and $C=2.99$ taken
from the fit of experimental data in Ref. \cite{Greilich_PRL06}. 

The frequency dependent regular part of electron $g$-factor in Eq.\eqref{g:omega0} does not provide by itself a complete description  of electron spin precession frequency dispersion connected with $g$-factor distribution.  This dispersion in a QD ensemble is  strongly affected by the QD shape and composition. The corresponding distribution of $g$-factors can be phenomenologically described by the Gaussian, $\rho_g(g_e)$, with the root mean square of electron $g$-factor distribution, $\Delta g_e$.

The dispersion of electron spin precession frequencies is affected also by fluctuations of hyperfine fields of nuclei that are collectively acting on the localized electron in a QD. The electron spin
precession frequency ${\bm \Omega}={\bm \Omega}_{{\rm L}}+ {\bm \omega}_{N} = \mu_B g_e {\bm B}/\hbar
+ {\bm \omega}_{N}$ contains a nuclear contribution, ${\bm
\omega_{N}}$, which is proportional to the projection of the
nuclear spin polarization on the external field (if external field
is much larger than nuclear field fluctuations, which usually is
the case)~\cite{Greilich_Science07}. The nuclear contribution is
connected with statistical fluctuations of the nuclear spin
polarization in a QD. The fluctuations are described by a Gaussian with the
dispersion $\Delta \omega_N$ proportional to $N^{-1/2}$, where $N$
is the number of nuclei in the QD volume~\cite{Merkulov_PRB02}. 
We  ignore the nuclear induced frequency focusing effect \cite{comment3}, which could modify the density of electron spin
precession mode  to a comb-like shape in a QD ensemble exposed to a pulse train excitation 
Ref. \cite{Greilich_Science07}. 

The resulting broadening of electron spin precession frequencies connected with $g$-factor dispersion and with nuclear fluctuations,  $\rho(\Omega)=\rho(\mu_B
g_eB/\hbar+\omega_{N})$ is also described by the Gaussian:
\begin{equation} \label{rho:g}
{\rho}(\Omega)=\frac{1}{{\sqrt{2\pi} \Delta \Omega }} \exp\left[ - \frac{(\Omega - \Omega_{{\rm L}}
)^2}{2 (\Delta \Omega)^2}\right],
\end{equation}
where $\Omega_{{\rm L}} (\omega_0)= \mu_B g_e({\omega}_0)B/\hbar
$ and $\Delta \Omega = \sqrt{(\mu_B \Delta g_e B / \hbar)^2 +
(\Delta \omega_{N})^2}$ is the total frequency dispersion.  
The range of electron spin precession frequencies generated by the pulse due to this dispersion   is shown in Fig.~\ref{fig:gfactor} 
by crosshatched region around the linear dependence
${\Omega}_{{\rm L}}(\omega_0)$. The calculations were conducted for $\Delta g_e = 0.0037$ \cite{Greilich_PRL06} and $\Delta\omega_{N}=0.37$~GHz extracted from the amplitude of random nuclear
fluctuation field of 7.5~mT 
\cite{petrov_prb}. We  assume  that $\Delta g_e$ and $\Delta
\omega_N$ is independent of the QD resonance energy. One can see that for used set of $\Delta g_e$ and $\Delta
\omega_N$ the dispersion of electron spin precession frequencies $\rho(\Omega)$ is much broader than one created  by the $g$-factor dependence on the excitation frequency (see insert in Fig.~\ref{fig:gfactor}). This additional broadening leads to the fast dephasing of electron spin polarization and should be taken into account in a description of the time dependence of FR  and ellipticity signals.

To obtain the time-dependent traces of the FR and
ellipticity signals  for the QD ensemble  we average
Eqs.~\eqref{farad:fin} and \eqref{ellipt:fin} over the distribution of
optical transition energies, $\rho_{\rm opt}(\omega_0)$, described by
Eq.~\eqref{rho:omega0} and over distribution of electron spin precession frequencies, $\rho(\Omega)$, described by Eq.~\eqref{rho:g}.  Without any calculations, however, one can notice from Eq.~\eqref{farad:fin} that in degenerate case when $\omega_{\rm pr}=\omega_{\mbox{}_{\rm P}}$, FR signal  vanishes if $S_z$ and $J_z$ and all dispersion functions are even functions of the detuning, $\omega_{\rm pr}-\omega_0$,   because the signal is proportional to the odd function of detuning $\Im[G(\omega_{\rm pr}-\omega_0)]$.  For degenerate case FR signal could arise for excitation at one side of $\rho_{\rm opt}(\omega_0)$ distribution, or  as a result of    dependence of electron $g$-factors or oscillator transition strengths  on the optical transition energy.

\subsection{Effects of a single pump pulse}

We start by considering the two color FR and ellipticity
signals excited by a single pump pulse as function of the time
delay between pump and probe pulses with frequencies
$\omega_{\mbox{}_{\mathrm P}}$ and $\omega_{\rm pr}$, respectively.  To 
 clarify qualitative differences between FR and ellipticity
signals  
we assume here and in the subsection \ref{subsect:inf}
that  $\rho_{\rm opt}(\omega_0)$ is
 independent of $\omega_0$ or, equivalently, that the
resonant excitation of QDs is performed at the maximum of
$\rho_{\rm opt}(\omega_0)$, which is so broad that $\Delta E\gg
\hbar/\tau_p$. To obtain nonvanishing FR signal, however, we take into account the  
dependence of an electron $g$-factor on the resonance transition
frequency, $\omega_0$, see Eq.~\eqref{g:omega0}.

\begin{figure}[hptb]
\includegraphics[width=\linewidth]{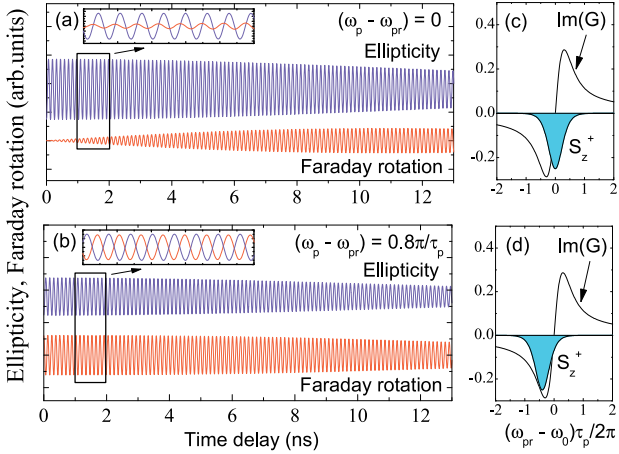}
\caption{Time resolved dependence of the pump-probe Faraday rotation and ellipticity signals initiated
in the QD ensemble by a single pump pulse for (a) the degenerate, $\omega_{\mbox{}_{\mathrm P}}=\omega_{\rm pr}$, and
(b) nondegenerate, $\omega_{\mbox{}_{\mathrm P}}\neq \omega_{\rm pr}$, regimes.
For clarity, the signals are vertically shifted from each other. Calculations are conducted for
the pump pulse with area $\Theta=\pi$ in the magnetic field $B=1$\,T by using $g$-factor spectral dependence described by Eq.~\eqref{g:omega0} and the parameters
$\tau_p=1.5$~ps, $\hbar\omega_{\rm pr}=1.4$~eV
and $\tau_{s,e}= 3$~$\mu{\mbox s}$ taken from Ref. \cite{Greilich_Science06}.
Insets show the Faraday rotation and ellipticity signals in a small range of delay times.
Faraday rotation curve in inset in panel (a) is multiplied by a factor of 2.
Panels (c) and (d) show $S_z^+$ and $\Im[G(\omega_{\rm pr}-\omega_0)]$ as a function of
$\omega_{\rm pr}-\omega_0$ for the degenerate and nondegenerate regimes, respectively.}
\label{fig:1pulse}
\end{figure}

Figure~\ref{fig:1pulse}(a) shows the traces of FR and ellipticity
signals for degenerate case ($\omega_{\mbox{}_{\mathrm P}}=\omega_{\rm pr}$) calculated in a magnetic field $B=1$~T for the QD ensemble with
the average $g$-factor, its dependence on $\omega_0$ and
$\tau_{s,e}$ extracted from the data of
Refs.~\cite{Greilich_PRL06,Greilich_Science06}. The trace of the
ellipticity signal demonstrates damped oscillations with the decay determined by the dispersion of electron spin
precession frequencies.  

 Figure~\ref{fig:1pulse}(a) shows also that the FR signal is absent at
zero delay time, as it was expected,  due to symmetric distribution $S_z$ and $\rho_{\rm opt}(\omega_0)$ around pumping frequency. Surprisingly, however, this signal  is growing in time. This happens because  the
electron spin distribution created by the pump pulse being
initially symmetric around $\omega_{\mbox{}_{\mathrm P}}$ (see~Fig.~\ref{fig:1pulse}(c)) is
gradually loosing its symmetry due to different Larmor precession
frequencies on the low- and high-energy wings of the QD
distribution as described by Eq.~\eqref{g:omega0}. This imbalance
of the electron spin polarization connected with electron
$g$-factor dependence on $\omega_0$ results in the growth of
the FR signal with time. The inset in
Fig.~\ref{fig:1pulse}(a) shows also  a phase shift between
the Faraday rotation and ellipticity signals. Calculations show (not presented) that oscillation
frequencies of the FR and ellipticity signals are also slightly different.  The effect is  connected with different
spectral contributions to the FR and ellipticity and results in a weak time dependence of the phase shift.

Figure~\ref{fig:1pulse}(b) shows FR and ellipticity signals
for the nondegenerate case where pump and probe pulses are detuned:
$\omega_{\mbox{}_{\mathrm P}}-\omega_{\rm pr}=0.8\pi/\tau_p$. The electron
spin polarization created in this case is not a symmetric function
of $\omega_{\rm pr} - \omega_0$ as one can see in
Fig.~\ref{fig:1pulse}(d), and the probe light measures the electron
spin polarization only at one of the spin distribution wings.
Therefore, at $t=0$ the FR signal is nonzero and its time dynamics is qualitatively
similar to that of the ellipticity signal.  It is seen from
inset in Fig.~\ref{fig:1pulse}(b) that the phase shift between
Faraday rotation and ellipticity signals is close to $\pi$. The
phase shift and the sign of the FR signal, correspondingly, depends on the sign of the pump-probe
detuning because $\Im[G(\omega_{\rm pr} -
\omega_0)]$ is an odd function of the detuning. 

\subsection{Effects of an infinite train of pump pulses}\label{subsect:inf}

\begin{figure}[hptb]
\includegraphics[width=\linewidth]{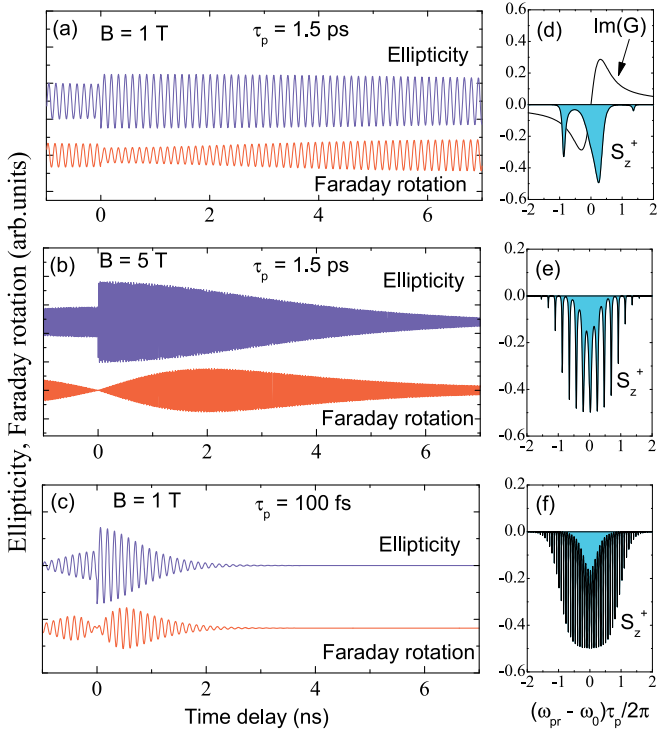}
\caption{Time resolved dependence of the single color pump-probe Faraday rotation and ellipticity signals
initiated in the QD ensemble by a train of pump pulses with the repetition period $T_R=13.2$\,ns and pulse duration $\tau_p=1.5$~ps ((a) and (b)) and $\tau_p=100$~fs (c), in the magnetic
fields $B=1$\,T (a) and (c), and $B=5$\,T (b). Calculations are conducted for pump pulses with the
area $\Theta=\pi$ using $g$-factor spectral dependence described by Eq.~\eqref{g:omega0} and the parameters $\hbar\omega_{\rm pr}=1.4$~eV and $\tau_{s,e}= 3$~$\mu{\mbox s}$
taken from Ref. \cite{Greilich_Science06}. Panels (d), (e) and (f) show the distributions $S_z^+(\omega_{\rm pr}-\omega_0)$
and $\Im[G(\omega_{\rm pr}-\omega_0)]$ created by the pico- and femtosecond pulse trains at $B=1$\,T and $B=5$\,T, respectively.}
\label{fig:train}
\end{figure}

Figure~\ref{fig:train} shows
the time dependent traces  of the FR
and ellipticity signals initiated by a train of short pulses of
circularly polarized light with the repetition period $T_R =
13.2$\,ns in degenerate regime $\omega_{\mbox{}_{\mathrm P}}=\omega_{\rm pr}$. Calculations were conducted for the pulse duration $\tau_p=1.5$~ps and magnetic fields  $B=1$\,T and  $B=5$\,T (panels (a) and (b), correspondingly). Panel (c) shows the results calculated for $\tau_p=100$~fs  and $B=1$\,T.
The commonly used repetition period of the mode-locked lasers $T_R$ is about 10~ns, which
is much shorter than the typical electron spin relaxation
time in a QD. As a result the infinite train of such pulses
creates a stationary distribution of rotating spin polarization
$S_z(\Omega;t)$,  which modifies strongly the FR and
ellipticity signals from those created by a single pump pulse. The traces in Figs.~\ref{fig:train}(a), (b) and
(c) are calculated by using the steady-state values
of the electron spin polarization defined by Eq.~\eqref{modelock}.
Here, like in the previous subsection, the dispersion of electron
spin precession frequency is described by the function
$g_e(\omega_0)$ given by Eq.~\eqref{g:omega0}.
Figures~\ref{fig:train}(d), (e) and (f) show the
electron spin distributions at the moment right after the pump
pulse arrival. One can see  that the latter
distributions are very different from those created by a single pulse and shown
in Figs.~\ref{fig:1pulse}(c) and \ref{fig:1pulse}(d).

In the case of excitation of QD ensemble by an infinite pump pulse train
 a steady state distribution of electron spin polarization is formed in accordance with Eq.~\eqref{modelock}. 
The modes satisfying the PSC: $\Omega = 2\pi N / T_{R}$ ($N$ is a large integer, $N\approx 100$ for $B=1$~T) provide an enhanced contribution to the electron spin polarization. The sum of these mode contributions to the spin polarization results in the constructive interference around the pulse arrival time due to the commensurability of the spin precession frequencies with the cyclic repetition  
frequency of the train, $2\pi/T_R$. 

The shape of the steady state distribution of electron spin polarization depends strongly 
on the number of precession modes, which satisfy PSC and, therefore, on pulse duration and on a magnetic field. At the relatively weak field, $B=1$~T, and pulse duration $\tau_p=1.5$~ps the distribution is asymmetric because only few modes satisfy PSC (see Fig.~\ref{fig:train}(d)).  In this case, the ellipticity and Faraday rotation signals are similar to each other analogously to the situation of detuned pump and probe, Fig.~\ref{fig:1pulse}(b).

There are  more modes satisfying PSC with increase of a magnetic field. This is because the dispersion of electron spin precession frequency increases linearly with magnetic field but the distance between the PSC  modes $2\pi/T_{R}$ does not change. The density of the mode satisfying the PSC increases and the steady state distribution of electron spin precession frequency at the moment of pulse arrival becomes more dense and symmetric (see Fig.~\ref{fig:train}(e), where the electron spin polarization was calculated for $B=5$~T). In this case the Faraday rotation signal and the ellipticity signal become phase shifted relative to each other. More importantly, the Faraday signal vanishes at zero delay between pump and probe pulses (Fig.~\ref{fig:train}(b)) similar to that for  the case the single pulse excitation with degenerate pump and probe pulses (c.f., Fig.~\ref{fig:1pulse}(a)). 
The shortening of the pulse duration ($\tau_p=100$~fs)  also leads to an increase of the number of modes satisfying PSC as it is clearly seen in Fig.~\ref{fig:1pulse}(f), because of the spectral width of the laser pulse increases with shortening of the pulse duration. The increase number of modes results in faster decay of ellipticity and FR rotation signals and in vanishing of the Faraday rotation signal at the moment of pulse arrival.
In two-color experiments (not shown) the ellipticity and the Faraday rotation signal time dependent traces become similar and the phase shift vanishes.

\subsection{{Effect of the pump and probe spectral position and the electron $g$-factor dispersion}}

\begin{figure}[hptb]
\includegraphics[width=\linewidth]{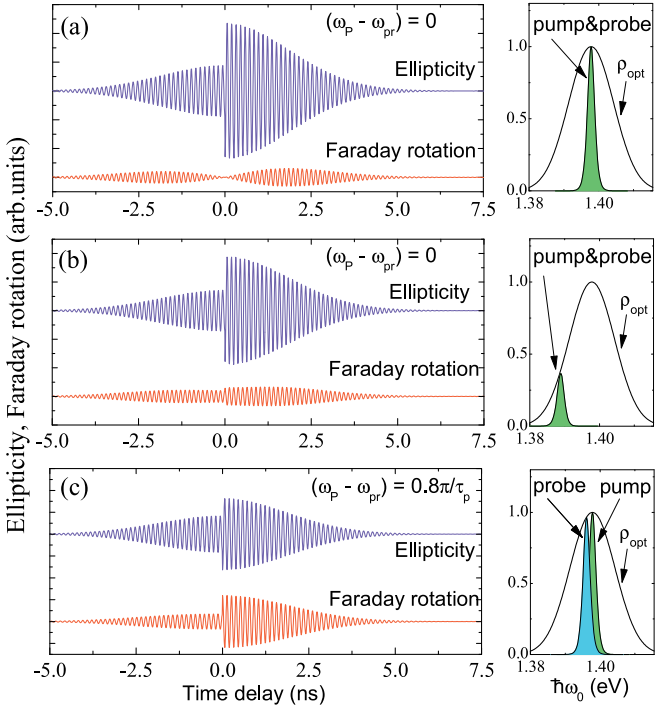}
\caption{Time resolved dependence of the pump-probe Faraday rotation and ellipticity signals initiated in
the QD ensemble by a train of pump pulses for the degenerate, $\omega_{\mathrm P}=\omega_{\rm pr}$, (panels (a)
and (b)) and nondegenerate, $\omega_{\mbox{}_{\mathrm P}}\neq \omega_{\rm pr}$, (panel (c)) regimes. In the right-hand side
panels we show the density of the optical transition energies  and the frequency
position of the pump and probe pulses. Calculations are conducted for pump pulses of the area $\Theta=\pi$, the magnetic
field $B=1$\,T, $\tau_p = 1.5$~ps, and $\tau_{s,e}= 3$~$\mu{\mbox s}$.}
\label{fig:spread}
\end{figure}

Now we turn to the effects of the spectral distribution of the QD
transition energies and of the distribution of  electron spin
precession frequencies described by Eqs.~\eqref{rho:omega0} and
\eqref{rho:g} on the time traces of the FR and ellipticity signals
created by the train of pulses with repetition period $T_R=13.2$\,
ns. Panels (a)-(c) in Fig.~\ref{fig:spread} show FR and ellipticity time-resolved dependences
for different spectral positions of pump and probe pulses which
are shown on the corresponding right-hand side panels. The panels
(a) and (b) show the traces for the same pump and probe carrying
frequencies (single color or degenerate pump-probe setup). This frequencies are tuned to the peak of QD
distribution in Fig.~\ref{fig:spread}(a) and to its
left wing in Fig.~\ref{fig:spread}(b). The panel (c) shows the
traces of the FR and ellipticity signals for the case when the
pump and probe pulse frequencies are in the vicinity of the
maximum of $\rho_{\rm opt}(\omega_0)$ but they are slightly detuned with
respect to each other. The distribution of optical transition
frequencies in the QD ensemble, $\rho_{\rm opt}(\omega_0)$, shown in
Fig.~\ref{fig:spread} by black curves is described by
Eq.~\eqref{rho:omega0}.
The calculation shows that the inclusion of additional dispersion
of electron spin precession frequencies described by
Eq.~\eqref{rho:g} leads to the faster decay of both the FR and
ellipticity signals due to a faster dephasing of electron spin
precession in the QD ensemble. It is worth noting that the FR
signal amplitude vanishes for degenerate pump and probe pulses
tuned to the maximum of the QD distribution, similarly to what was
experimentally observed in Ref.~\cite{Greilich_pss}. The nonzero
FR signal at zero delay time seen in Figs.~\ref{fig:spread}(b) and
\ref{fig:spread}(c) is related to the asymmetry in the QD density
distribution which is revealed in the case of pump pulse detuned
from the $\bar{\omega}_{0}$ ($\hbar\bar{\omega}_0 = 1.4$~eV). In the case of two color experiments,
traces of the pump-probe FR and ellipticity signals are very
similar as it is seen in Fig.~\ref{fig:spread}(c).

\begin{figure}[hptb]
\includegraphics[width=0.8\linewidth]{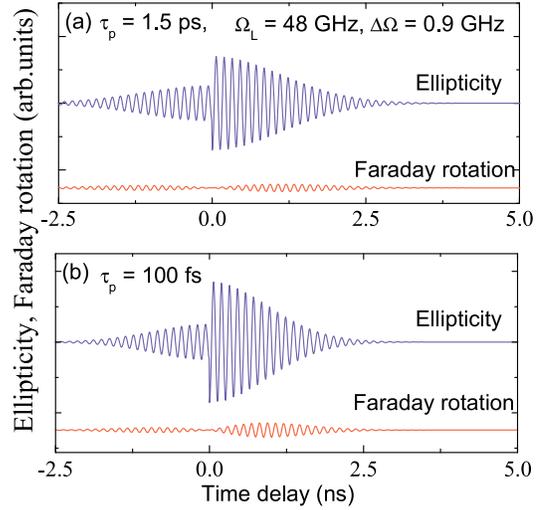}
\caption{Time resolved dependences of the degenerated pump-probe Faraday rotation and ellipticity signals in inhomogeneous ensemble of QDs excited by trains of pulses with duration 1.5~ps (panel (a)) and 100~fs (panel (b)), correspondingly. Calculations are conducted for the set of parameters the same as in
Figs.~\ref{fig:train}(a) and \ref{fig:train}(c). Note, that the time scale used here is different from the time scale used in Fig.~\ref{fig:train}.}
\label{fig:ps_vs_fs}
\end{figure}

\begin{figure*}[hptb]
\includegraphics[width=0.7\linewidth]{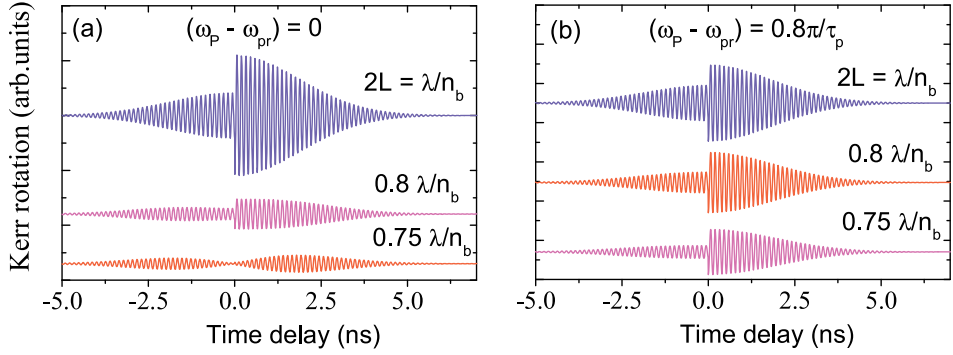}
\caption{Time resolved dependence of pump-probe Kerr rotation signals initiated in the
QD ensemble by a train of pump pulses for the degenerate (panel (a))
and nondegenerate (panel (b)) regimes. The traces are calculated for three thicknesses of the cap layer
$2L = \lambda/n_b,~0.8\lambda/n_b,~{\rm and}~0.75\lambda/n_b$, where $n_b \equiv \sqrt{\varepsilon_b}$. Calculations are conducted for the set of parameters the same as in
Figs.~\ref{fig:spread}(a) and \ref{fig:spread}(c). }
\label{fig:kerr}
\end{figure*}

Figure~\ref{fig:ps_vs_fs} compares the Faraday rotation and ellipticity signals for laser trains with different pulse duration: 1.5~ps (a) and 100~fs (b).  Calculations  that take into account spectral distribution of optical transition energies in the QD ensemble $\rho_{\rm opt}(\omega_0)$ and total Larmor frequency dispersion where conducted in degenerate regime for resonant excitation of QDs at the maximum of $\rho_{\rm opt}(\omega_0)$ ($\omega_{\mbox{}_{\rm P}}=\omega_{\rm pr}=\omega_0$). The case of ps-excitation differs from one demonstrated in Fig.~\ref{fig:spread}(a) only by larger dispersion of electron spin precession modes $\Delta \Omega=0.9$~GHz (time dependent traces in Fig. ~\ref{fig:spread}(a) were calculated for $\Delta\Omega=0.5$~GHz) and, therefore, they show faster decay of the signals. The spectral width of pulses in the fs-pulse train is larger than the distribution $\rho_{\rm opt}(\omega_0)$ used in this calculation. This effectively decreases the  number of electron spin precession modes contributing to the FR and ellipticity signals  explaining dephasing decay, which is slightly weaker than the decay shown in Figure~\ref{fig:train}(c). 

Surprisingly, Fig.~\ref{fig:ps_vs_fs} shows that signals created by  ps- and fs-pulse trains are very similar and the signal decays are almost the same. This could occur only if numbers of electron precession modes satisfying the PSC for both excitations are comparable. In the case of ps-train excitation the dispersion of electron spin precession modes is controlled by $\Delta g$ and $\Delta \omega_N$ connected with the shape and composition fluctuations of QDs  and the nuclear field fluctuations, correspondingly.  Due to the small spectral width of the ps-excitation only a small part of the electron spin precession mode dispersion is determined by the frequency dependence of the electron $g$-factor.  This is not the case, however, for fs-pulse train excitation, where the dispersion of electron spin precession modes have significant contribution connected with frequency dependence of an average $g$-factor $g(\omega_0)$ on the spectral width of the fs-pulses. %\addMisha{Hence, for a ps-excitation the dispersion of spin precession frequencies is determined mostly by the nuclear field fluctuations while for a fs-excitation by the optical frequency dependence of the $g$-factor.}

Finally, Fig.~\ref{fig:kerr} shows traces of the KR signal
calculated by using Eq.~\eqref{kerr:fin} for three different
thicknesses of the cap layer and the same parameters as in
Fig.~\ref{fig:spread}. One can see that, for particular cap layer
thicknesses, the KR signal looks like the either FR or
ellipticity signals. However, in general the trace of the KR
signal is a linear combination the FR and ellipticity signals with
their partial contributions depending on the cap layer thickness.

Note that $L = \lambda/2n_b$ (top curves in Fig.~\ref{fig:kerr}) corresponds to the real thickness $L = 115$~nm of cap layer of the QD
sample which was investigated in Ref.~\cite{Greilich_Science06}. For this cap layer time dependent trace of the KR signal  is similar to that of the ellipticity signal.

\section{Summary}
The formalism presented here provides a complete theoretical
description of single- and two-color pump-probe Faraday or Kerr
rotation and ellipticity experiments in an ensemble of singly
charged QDs. The analytical expressions describing the electron
spin polarization created by a circularly polarized pump pulse or
by a train of such pulses are derived. The expressions for the
magnitudes of the Faraday, Kerr and ellipticity signals are
presented. 

The developed theory shows that the pump-probe Faraday
rotation and ellipticity experiments measure the electron spin
precession in slightly different subsets of QDs of the ensemble
leading to the different oscillation frequencies and shapes of the
corresponding time-dependent traces. The time-dependent traces of
the pump-probe Kerr rotation signal are linear superpositions of
the Faraday rotation and ellipticity signals whose relative
weights depend on the cap layer thickness. 

The modeling of
time-dependent traces of the Faraday rotation signal shows their
high sensitivity to the inhomogeneous properties of the QD
ensemble, such as the transition-frequency dependence of electron
$g$-factor and the nuclear-induced dispersion,  as well as to the
excitation conditions, such as pump and probe pulse detuning,
single pulse versus train of pulses excitation, and the pumping
intensity. The pump-probe Faraday and Kerr rotation and
ellipticity experiments can provide a complementary information
about inhomogeneous properties of QD ensembles.

\section*{ACKNOWLEDGMENTS}
The authors thank M. Bayer and D. R. Yakovlev  for the encouraging
discussions and support as well as hospitality at the TU
 Dortmund University,   I. V. Ignatiev and S. Carter for useful comments on the
manuscript.  I. A. Y., M. M. G. and E. L. I. acknowledge the
financial support from RFBR, Programmes of RAS and  the Deutsche
Forschungsgemeinschaft (SPP1285). A. L. E. acknowledges support of
the Office of Naval Research and Alexander-von-Humboldt
Foundation. M. M. G. is grateful to the ``Dynasty'' Foundation ---
ICFPM.


\begin{thebibliography}{55}
\bibitem{Onecomponent} Here we consider only one term, which is proportional to $\mathrm
e^{-\mathrm i \omega_{\rm pr} t}$, for the complex conjugate term the calculation is analogous.
\bibitem{comment3}  These
modifications have no general nature: they depend on many different
parameters, particularly, on details of sample optical excitation and should be considered
separately for each QD ensemble. In addition the modification of
the electron spin precession frequency by the nuclei requires some
time and should not affect the FR, KR and ellipticity signals
during long times of nuclear spin dynamics. The light induced nuclear
polarization  also can be eliminated by the microwave field.
\end{thebibliography}
\end{document}